\documentclass[useAMS,usenatbib]{mn2e}
\usepackage{epsfig}
\usepackage{pgf}
\def\lesssim{\la}

\title[]{
Structure in phase space associated with spiral and bar density waves
in an N-body hybrid galactic disk
}

\author[]{
Alice C. Quillen$^1$,  
Jamie Dougherty$^1$, 
Micaela B. Bagley$^{1,2}$, 
\newauthor
Ivan Minchev$^{3,4}$,  \&
Justin Comparetta$^1$  \\
$^1$ Department of Physics and Astronomy, University of Rochester, Rochester, NY 14627, USA; aquillen@pas.rochester.edu \\
$^2$ Steward Observatory, 933 N. Cherry St., University of Arizona, 
AZ 85721, USA \\
$^3$ Observatoire Astronomique, Universit\'e de Strasbourg, CNRS UMR 7550, 67000 Strasbourg, France \\
$^4$ Astrophysikalisches Institut Potsdam, An der Sternwarte 16,
D-14482, Potsdam, Germany
}

\begin{document}

\label{firstpage}
\maketitle

\begin{abstract}
An N-body hybrid simulation, integrating both massive
and tracer particles, of a Galactic disk 
is used to study the stellar phase space distribution or
velocity distributions in different local neighborhoods.
Pattern speeds identified in Fourier spectrograms 
suggest that two-armed and three-armed spiral density waves, 
a bar and a lopsided motion are coupled in this simulation,  
with resonances of one pattern lying near resonances of other patterns.
We construct radial and tangential ($uv$) velocity distributions  
from particles in different local neighborhoods.
More than one clump is common in these local velocity distributions
regardless of the position in the disk. 
Features in the velocity distribution 
observed at one galactic radius are also seen in nearby 
neighborhoods (at larger and smaller radii) but with shifted mean $v$ values.  
This is expected if the $v$ velocity component of a clump sets 
the mean orbital galactic radius of its stars.
We find that gaps in the velocity distribution 
are associated with the radii of kinks or discontinuities in the spiral arms. 
These gaps also seem to be associated with Lindblad resonances 
with spiral density waves and so denote boundaries between
different dominant patterns in the disk. 
We discuss implications for interpretations of the Milky Way disk 
based on local velocity distributions.
Velocity distributions created from regions just
outside the bar's Outer Lindblad resonance
and with the bar oriented at $45^\circ$ from the Sun-Galactic center
line more closely resemble that seen in the solar neighborhood (triangular
in shape at lower $uv$ and with a Hercules like stream)
when there is a strong nearby spiral arm,  
consistent with the observed Centaurus Arm tangent,
just interior to the solar neighborhood.

\end{abstract}

\section{Introduction}

The velocity distribution of stars in the solar neighborhood contains
structure that has been particularly clearly
revealed from Hipparcos observations
\citep{dehnen98,famaey04}.
Much of this structure was previously associated with moving
groups \citep{eggen}.  Moving groups are associations of
stars that are kinematically similar or have similar space motions.
Young, early-type stars can be moving together because they carry the kinematic
signature of their birth \citep{eggen}.  However, a number of the kinematic
clumps identified through the kinematic studies also contain
later type and older stars \citep{dehnen98,famaey08,nordstrom04,arifyanto06}.

In a small region or neighborhood of a galaxy, 
structure in the stellar velocity distribution
can be caused by different dynamical processes.  
Bar or spiral patterns can induce structure in the phase space distribution, 
particularly in the vicinity of Lindblad resonances 
\citep{yuan97,dehnen99,fux01,quillen05,minchev06,antoja09,gardner10,lepine10}.
Mergers and orbiting satellite subhalos can leave behind stellar streams
(e.g., \citealt{bekki03,meza05,helmi06,gomez10}).  
In a local neighborhood, these could be seen as 
increases in the stellar number density at a particular velocity.
%
%

The velocity vector of a star in the solar neighborhood
can be described in Galactocentric coordinates with components
$(u,v,w)$. Here $u$ is the radial
component with positive $u$ for motion toward the Galactic center.
The $v$ component is the tangential velocity component 
subtracted by the velocity of a particle in a circular orbit 
computed from the azimuthally averaged midplane gravitational potential. 
The $v$ component is positive for a particle
with tangential component larger than the circular velocity that is
moving in the same direction as disk rotation.
The $w$ component is the velocity in the $z$ direction.

\citet{quillen05} showed that Lindblad resonances associated 
with a spiral pattern, given by their approximate
$v$ velocity component, were likely locations for
deficits in the velocity distribution in the solar neighborhood
(also see \citealt{fux01,lepine10}).
On either side of the resonance the shape of orbits changes. For example
closed or periodic orbits can shift orientation so they
are parallel to a bar on one side of the resonance
and perpendicular on the other side. On resonance there may be no nearly
circular orbits. As a consequence in a specific local neighborhood 
there can be a particular velocity associated with resonance 
that corresponds only to 
orbits with high epicyclic variations.  These may not be well populated
by stars leading to a deficit of stars in the local velocity distribution
at this particular velocity.
This picture 
is also consistent with the division between the Hercules
stream and the thin disk stars that is likely caused by the $m=2$ outer Lindblad
resonance (OLR) with the Galactic bar \citep{dehnen00,minchev07,gardner10}. 

Structure seen in the velocity distribution can also be caused by
recent large scale perturbations.
Perturbations on the Galactic disk caused by bar formation or a galactic
encounter, can perturb the action angle distribution.  
Subsequent phase wrapping can induce structure in the velocity distribution
\citep{minchev09,quillen09}.
Transient spiral density waves can also leave behind structure
in the local phase space distribution \citep{desimone04}.
The local velocity distributions can be particularly complex when both
spiral and bar perturbations are present or when multiple spiral patterns
are present
\citep{quillen03,minchev06,dalia08,antoja09,minchev10a,minchev10b,minchev11}.

The variety of explanations for structure in the regional phase 
space distribution presents a challenge for interpretation of 
current and forthcoming radial velocity and proper motion surveys.
We may better understand the imprint of different dynamical structures
in the phase space distribution by probing simplified numerical models.
Most numerical studies of structure in localized regions
of the galactic disk phase space 
distribution has been done using test particle integrations
(e.g., \citealt{dehnen00,desimone04,quillen05,dalia08,antoja09,minchev09,quillen09,gardner10}).
Few studies have probed the local phase space distribution 
in an N-body simulation of a galactic disk (but see \citealt{fux00}) 
because of the large number of particles
required to resolve structure both in physical and velocity 
space.  In this paper we use an N-body hybrid simulation,
integrating both massive and tracer particles, that exhibits both
spiral and bar structures to 
probe the velocity distribution in different regions of a simulated
disk galaxy.
Our goal is to relate structure in the local phase space distribution
to bar and spiral patterns that we can directly measure in the simulation.

\section{Simulations}

In this section we describe our N-body hybrid simulations and their initial conditions.
We also describe our procedure for making spectrograms of the Fourier 
components so that we can identify pattern speeds in the simulations.

\subsection{The N-body hybrid code}

The N-body integrator used is a direct-summation code
called $\phi$-GRAPE  \citep{harfst07} that employs a 4-th order 
Hermite integration
scheme with hierarchical commensurate block times steps \citep{makino92}.
Instead of using special purpose GRAPE hardware we use 
the Sapporo subroutine library \citep{gaburov09} that closely
matches the GRAPE-6 subroutine library \citep{makino03} but
allows the force computations to be done
on graphics processing units (GPUs).
The integrator has been modified so that massless tracer particles
can simultaneously be integrated along with the massive
particles \citep{comparetta10}.  
The tracer particles help us resolve structure in phase space 
without compromising the capability to carry out a self-consistent
N-body simulation.  The simulations were run on a node
with 2 NVIDIA Quadro FX 5800 GPUs. The cards are capable
of performing double precision floating point computations,
however the computations were optimized on these 
GPUs to achieve double precision accuracy with
single precision computations using a corrector 
\citep{gaburov09}.  The Quadro FX cards were designed to target the Computer-Aided Design and Digital Content Creation audience and so were slower but more accurate than other similar
video cards even though they did not impliment error correction code.

\subsection{Initial Conditions for the Integration}

The initial conditions for a model
Milky Way galaxy were made with numerical phase phase
distribution functions using the method discussed by
\citet{widrow08} and their numerical routines which
are described by \citet{widrow08,kuijken95,widrow05}.
This code, called GalacticICS, 
computes a gravitational potential for bulge, disk and halo
components, then computes the distribution function for each component.
Particle initial conditions are then computed for each component.
The galactic bulge is consistent with a Sercic law for the projected density.  
The halo density profile is described by 5 parameters, a halo scale
length, a cusp exponent, a velocity dispersion, and an inner truncation radius
and truncation smoothness, so it is more general than an NFW profile.
The disk falls off exponentially with radius and as a sech$^2$ with 
vertical height.
We have adopted the parameters for the Milky Way model 
listed in Table 2 by \citet{widrow08}.
A suite of observational constraints are used to constrain the parameters
of this model, including 
the Oort constants, the bulge dispersion, the local
velocity ellipsoid, the outer rotation curve, vertical forces
above the disk and the disk surface brightness profile.

The number of massive particles we simulated
for the halo, bulge and disk are 50,000, 150,000 and 800,000, respectively.
The halo is live.  The number of test (massless) particles in the disk
is 3 million and exceeds the number of massive particles by a factor of 3.  
The total mass in disk, bulge and halo is 
$5.3 \times 10^{10}$, $8.3 \times 10^9$, and $4.6 \times 10^{11} M_\odot, $
respectively.
The mass of halo, bulge and disk particles is 
$9.2 \times 10^6, 5.5\times 10^4, $ and $6.6 \times 10^4 M_\odot$, respectively.
Snapshots were output every 5 Myrs and our simulation ran for 1.3 Gyr.

The smoothing length is 10 pc and is similar to the mean disk interparticle
spacing. We have checked that the center of mass of the simulation drifts
a distance shorter than the smoothing length during the entire simulation.
The parameters used to
choose timesteps in the Hermite integrator were $\eta = 0.1$ and 
$\eta_s = 0.01$ (see \citealt{makino92,harfst07}).  
The minimum timestep is proportional to $\sqrt{\eta}$ and is
estimated from the ratio of the acceleration to the jerk.
Our value places the minimum possible timestep
near the boundary of recommended practice\footnote{N-BODY 
Simulation Techniques: There's a right way and a wrong way,
by Katz, N., http://supernova.lbl.gov/$\sim$evlinder/umass/sumold/com8lx.ps}.
The total energy drops by 0.2\% during the simulation, 
at the boundary of recommended practice that suggests a maximum 
energy error\footnote{http://www.ifa.hawaii.edu/$\sim$barnes/ast626\_09/NBody.pdf}
$\delta E/E \lesssim 1/\sqrt{N}$ (see \citealt{quinlan92} on how the shadow distance 
depends on noise and softening).

In our goal to well resolve the galactic disk we have purposely undersampled
both halo and bulge and so have introduced spurious numerical noise
into the simulation from these populations.
 The product of the mass of halo particles and the halo density
density exceeds that of the same product for the disk implying
that numerical heating from the massive 
particles in the disk is lower than that induced by 
the live halo particles. The large timesteps, relatively small smoothing length and large halo particles
imply that this simulation is noisier than many discussed
in the literature.

We first discuss the morphology of the simulated galaxy and then
use the spectrograms constructed from the disk density as a function
of time to identify patterns.

\section{Morphology}

\begin{figure*}  
\includegraphics[angle=0,width=5in]{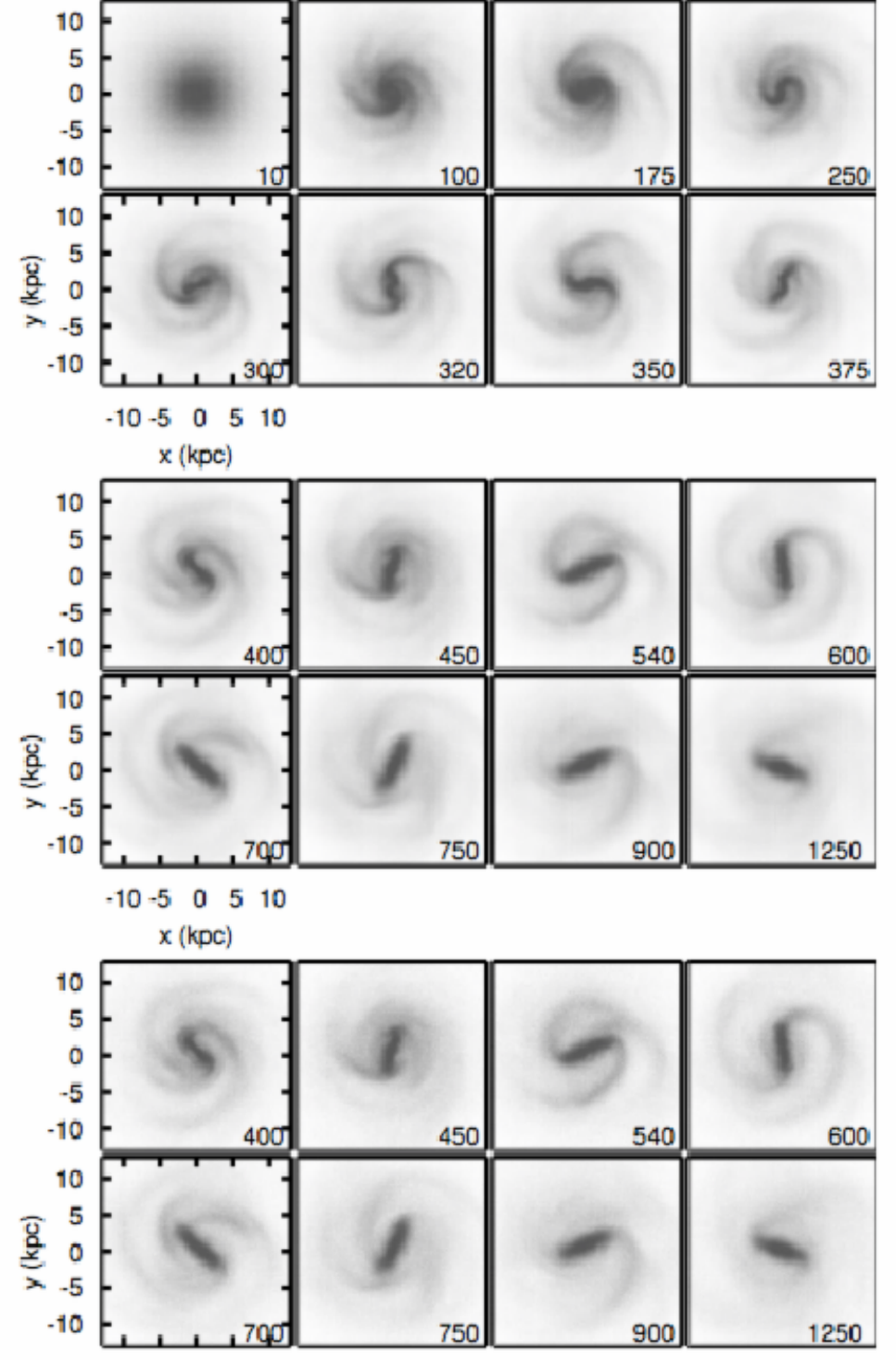} \\
\begin{center}
$\begin{array}{c}
%
\end{array}$
\end{center}
\caption{
Disk stellar densities projected into the $xy$ plane.
during early spiral arm and bar growth.
The time of each snapshot is shown in Myr on the lower right of each panel.
a) Here we show the earlier part of the simulation. 
During this time we see initial stochastic 
spiral arm growth followed by the onset of bar instability.
b) Here we show snapshots from the later part of the simulation.
We can see that the bar is not symmetrical; it can be lopsided.
At times the galaxy appears to have three-arms (e.g., at $T=750$ Myr).
Kinks and discontinuities are commonly seen in the spiral structure.
c) The same as b) except only massive disk particles are shown.
There is no significant difference between the distributions
of massive and massless disk particles.
\label{fig:dens}
}
\end{figure*}

In Figures \ref{fig:dens} we show
the disk density projected onto the midplane in Cartesian coordinates
for a few of the simulation outputs (snapshots).  
Disk particles are used to create these density histograms,
but bulge and halo particles are neglected.  
In Figures \ref{fig:dens} a,b both massive and tracer particles are shown.
Figure \ref{fig:dens}c is equivalent to Figure \ref{fig:dens}b except
only massive disk particles are shown.  We find no significant differences between
the distribution of massive and tracer disk particles,
consistent with the previous study using the same
hybrid code \citep{comparetta10}.

The numbers of massive particles is sufficiently low that
numerical noise is present and amplified to become spiral structure.
During the initial 40 snapshots (over a total time of 200 Myr) 
(shown in Figure \ref{fig:dens}a)
two-armed and three-armed spiral armed waves
grow in the region with radius 5--12 kpc.  
At around a time $T\sim 225$ Myr an open and somewhat lopsided 
axisymmetric structure grows within a radius of 4~kpc that
becomes the bar.    
At later times both two-armed and three-armed structure exists
outside the bar (shown in Figure \ref{fig:dens}b).

As our halo is live (and not
fixed) lopsided motions are not prevented in the simulation.
Lopsided motions and three-armed structure has previously
been observed in N-body simulations of isolated disks 
(e.g, see the simulations by \citealt{chilingarian}).
The bulge does not remain fixed during the simulation as
a lopsided perturbation develops that moves its center
slowly up to a distance of 0.5 kpc away
from the initial origin.  This distance is larger than the
distance our center of mass drifts during the simulation 
(less than 10 pc).
The lopsided motion may be related to the development of three-armed structures in the 
disk and asymmetries in the bar.    
Lopsidedness is common in spiral galaxies (see \citealt{jog09} for a review), 
and can spontaneously grow in N-body simulations of disks 
\citep{revaz04,saha07}.
If the pattern speed is slow, a lopsided mode 
can be long lasting \citep{ideta02}.
There is no single strong frequency or amplitude
associated with the lopsided motion in our simulation, 
though the motion is primarily counter-clockwise 
and moving with the direction of galactic rotation.
We have checked that the center of mass of the simulation drifts less than 
the smoothing length of 10 pc
so the lopsided motion is not due to an unphysical drift in the total momentum
in the simulation.
However, the growth of the lopsided motion could be due to noise
associated with our small smoothing length and large halo mass particles. 
Simulations of a disk residing in a rigid halo would
not exhibit lopsided motions.
However as most galaxies display lopsidedness we can consider
it a realistic characteristic of a galactic disk.
We will discuss this motion later on when we discuss the $m=1$ Fourier component spectrograms.

\begin{figure*}  
\includegraphics[angle=0,width=7.0in]{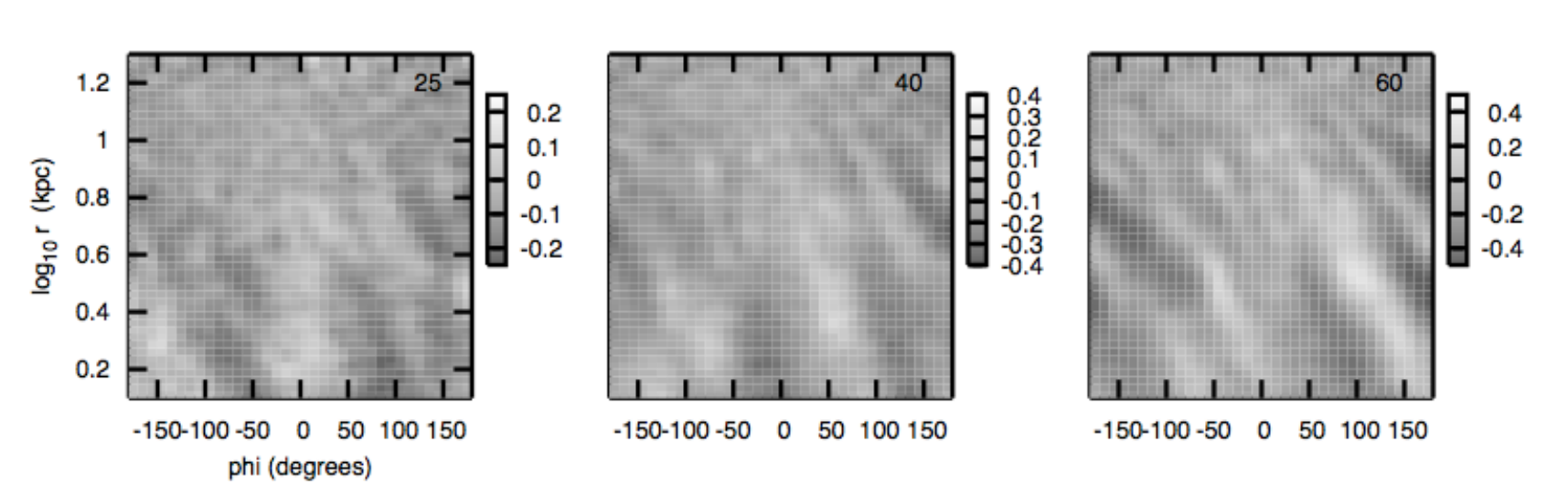}
\caption{Differential surface density 
as a function of the log of the radius $r$ 
and azimuthal angle $\phi$.
Here we show the surface density at three different times 
early in the simulation and during spiral arm growth.   
The times of the snapshots in Myr are shown in the upper right hand corners 
of each plot.  Patterns move to the right with time and trailing structures have
negative slopes.  A logarithmic spiral arm has high densities along a line on this plot 
with pitch angle given by the slope of the line.
The vertical feature in the leftmost
panel tilts and moves to the right in the center and then rightmost panel.  
In the rightmost panel
it is a stronger perturbation and with a larger and trailing pitch angle. 
The spiral arm shown in the center of the leftmost panel
becomes more tightly wound, 
and stronger, as would be expected from a swing amplification mechanism.
\label{fig:l1}
}
\end{figure*}

\begin{figure*}  
\includegraphics[angle=0,width=7.0in]{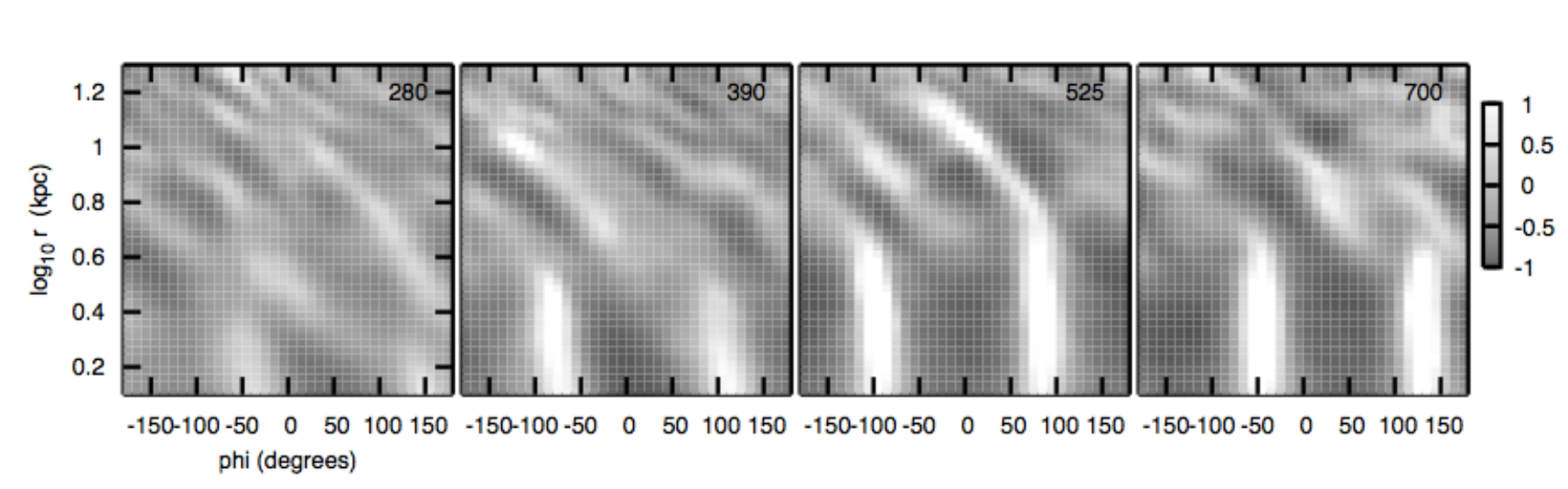}
\caption{Similar to Figure \ref{fig:l1} except showing the differential surface
density after bar formation.  On the lower parts of these panels the vertical
features are the bar.  There are two vertical lines per panel 
as would be expected from an $m=2$ bar-like structure.  
Exterior to the bar, both two-armed and three-armed patterns
are seen and their patterns move more slowly than the bar.
On an animated version of this plot it is clear that patterns in the inner 
galaxy are faster than those in the outer galaxy.
Interference between the patterns sometimes causes armlets to appear and disappear.
\label{fig:l2}
}
\end{figure*}

We also show the disk density in polar coordinates in Figures \ref{fig:l1} and
\ref{fig:l2}. 
These figures 
show differential density histograms as a function
of radius and azimuthal angle ($r,\phi$) (or in cylindrical coordinates
again projected onto the midplane).
The bins to create these two plots are logarithmically spaced in radius.
The disk density $\Sigma(r,\phi)$ 
in these two plots has been normalized at each radius. 
These histogram show $(\Sigma (r,\phi) - \bar \Sigma(r)) \bar \Sigma(r)^{-1}$
where $\bar \Sigma(r)$ is the mean density at radius $r$
averaged over the angle $\phi$.
A logarithmic spiral would give high densities along a straight line
in these plots, with slope set by the spiral arm pitch angle.
Two armed structures would correspond to two linear but sloped features, each
separated by 180$^\circ$.
Trailing spiral structures have negative slopes.
A bar on these diagrams would correspond to two vertical features
separated by 180$^\circ$.
The centroid of the galaxy bulge is subtracted prior to computing
the density histograms shown in \ref{fig:l1} and \ref{fig:l2} 
in polar coordinates but not in Figure \ref{fig:dens} 
that is in Cartesian coordinates.  
The times of the snapshots in Myr are shown in the upper right hand corners 
of each panel shown in Figures \ref{fig:dens} -- \ref{fig:l2}.  

In Figure \ref{fig:l1}
 the differential surface density in polar coordinates is shown
at three different times 
early in the simulation and during the early phase of spiral arm growth.   
The nearly vertical strip in the leftmost
panel tilts and moves to the right at later times.  In the right most panel
it is a stronger perturbation and with a larger, trailing pitch angle. 
The spiral arm shown in the center of the leftmost panel
becomes more tightly wound, and stronger (as seen on the right) 
as would be expected from a swing amplification mechanism.
After about 50 Myr the spiral patterns
are more slowly evolving.  Other than during
bar formation at about 330 Myr, 
and in the beginning of the simulation before 50 Myr, 
the spiral pitch angles do not seem to vary significantly. 

In Figure \ref{fig:l2}  the differential surface density is also shown at three different
times but after bar formation.  
On the lower parts of the panels the 
bar is seen as pair of vertical
strips.  We can see that the bar length increases later in the simulation.
Exterior to the bar, both two-armed and three-armed patterns
are seen and their patterns move more slowly than the bar.
The winding or pitch angle of spiral features
is about 24$^\circ$ and $\alpha \equiv d\phi/d\ln r \sim 2.3$. 
This pitch angle exceeds many but not all estimates for spiral arm
pitch angles near the solar neighborhood (see Table 1 by \citealt{vallee08}; most
pitch angle estimates are $\sim 13^\circ$).
Other than during bar formation and before 50 Myr, the spiral pitch angles do
 not seem to vary significantly.  
  
When animated the two-armed and three-armed features appear to constructively 
and destructively add with density peaks occurring at times and positions 
when two patterns lie on top of one another
(e.g., as discussed by \citealt{henry03,meidt09}).
The sum of two different patterns can cause individual
arm peaks to shift position and temporarily appear more or less tightly wound.

Kinks or changes in pitch angle of the spiral arms 
seen in the density images in Cartesian coordinates 
(in Figure \ref{fig:dens}b) correspond to places where the spiral arms
are discontinuous or change slope in the density distribution
as seen in the polar plots (e.g., Figure \ref{fig:l2}).
For example, the spiral pattern just outside the bar has a 
lower pattern speed than the bar. Consequently the phase of the spiral arms
varies with respect to the bar. At times spiral arms appear to 
be linked with the bar and at other times they appear separated from
the bar. Spiral arms in the outer galaxy have slower pattern
speeds than those in the inner galaxy. At times the spiral arms appear
to be connected and at other times, armlets or kinks are seen.

\section{Spiral and Bar Patterns in the Simulation}

Before we discuss velocity distributions extracted from local neighborhoods we first discuss patterns
of waves measured from the simulations.

\subsection{Constructing Spectrograms}

We constructed spectrograms using the procedure 
outlined by \citet{meidt08} in their section 3.2 
(also see section 3.4 by \citealt{masset97}). 
In each radial bin, defined by a binning function $b(r)$, 
and at each snapshot time $t$ we compute
\begin{eqnarray}
W_C^m (r,t) &=& \sum_i \cos(m \theta_i) b(r) \nonumber \\
W_S^m (r,t) &=& \sum_i \sin(m \theta_i) b(r) \label{eqn:wcs}
\end{eqnarray}
where the sum is over all disk particles in the radial bin (both
massive and tracer) and the integer, $m$, is the azimuthal wave number.  
Bulge and halo particles are ignored.
We use logarithmically spaced radial bins.
We compute the Fourier transform of the complex function 
composed of $W_C^m(r,t)$ and $W_S^m(r,t)$ or
\begin{equation}
\tilde W^m (\omega,r) = \int_{T_1}^{T_2} 
e^{i\omega t} h(t)\left[W_C^m(r,t) + i W_S^m(r,t)\right] dt
\end{equation}
where $h(t)$ is a Hanning function spanning our time window $[T_1,T_2]$
between snapshot time $T_1$ and $T_2$.
In our spectrogram figures we show the amplitude 
of the complex Fourier components 
or $|\tilde W^m(\omega,r)|$.  
In our spectrograms the angular frequency, $\omega$, increases along the 
$y$-axis and $\log_{10} r$ increases along the $x$-axis).

A bar or two-armed
spiral structure has strong $m=2$ Fourier components in its
density distribution (calculated with
equation \ref{eqn:wcs}).
The pattern speed of the wave is related to
its angular frequency by $\Omega_p \equiv \omega/m$.
Pattern speeds are found by dividing the angular frequencies of 
horizontal features in the spectrograms by the integer $m$.

During an N-body simulation with a non-stationary or live halo, 
lopsided modes or waves may develop and the bulge and
central disk of the galaxy may not remain at a fixed position.
If the Fourier components are measured respect to a radial position
that is not centered on the bulge centroid then an $m=1$ component
can generate other Fourier components.  This is sometimes called `aliasing' or `mode-mixing' 
(e.g., \citealt{olling03}).
As a consequence we subtract the position
of the centroid of the galaxy bulge prior to computing
the Fourier coefficients with $m>1$.  We will also discuss
the $m=1$ spectrogram computed without subtracting the bulge centroid position.

\subsection{Pattern speeds}

\begin{figure*}  
\begin{center}
$\begin{array}{cc}
\includegraphics[width=3.5in]{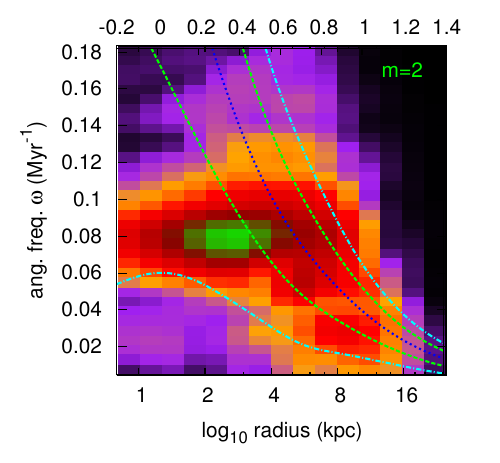} &
\includegraphics[width=3.5in]{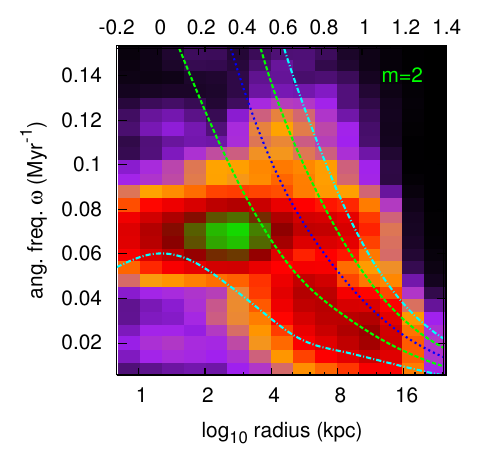}
\end{array}$
\end{center}
\caption{
Spectrogram of the $m=2$ Fourier components in a window spanning 
a) from time $T=350$ to 950 Myr and 
b) from time $T=750$ to 1305 Myr. 
The $y$ axis is angular frequency in Myr$^{-1}$ 
and the $x$ axis is radius in kpc
but shown on a $\log_{10}$ scale.  
Overplotted are five lines that are twice the following angular frequencies
in order of bottom to top
$\Omega - \kappa/2$, $\Omega - \kappa/4$, $\Omega$, $\Omega + \kappa/4$, 
and $\Omega + \kappa/2$.
The bar, once present, is stable
and has a slowly decreasing pattern speed.   
There are at least two two-armed spiral patterns.
The spiral patterns at both early and later times seem to extend between
their inner $m=2$ or 4 Lindblad resonance out to their corotation radii. 
\label{fig:spec2}
}
\end{figure*}

\begin{figure*}  
\begin{center}
$\begin{array}{cc}
\includegraphics[angle=0,width=3.5in]{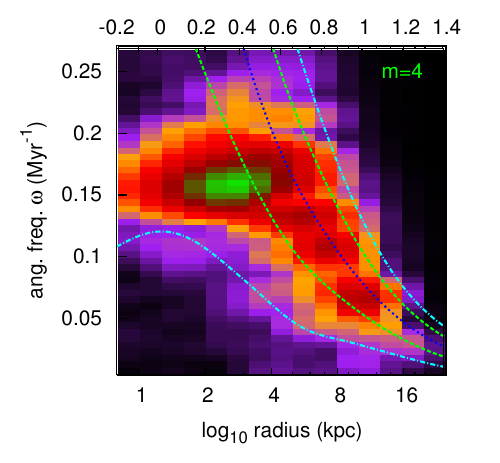} &
\includegraphics[angle=0,width=3.5in]{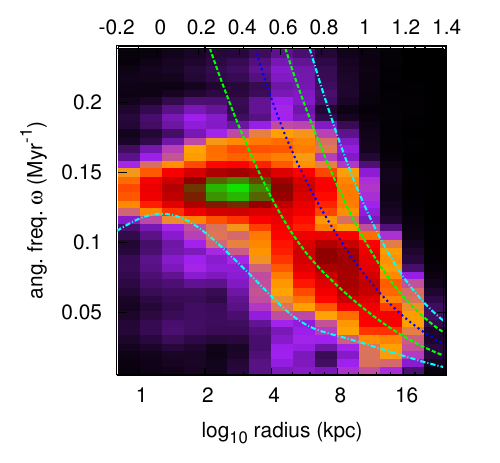}
\end{array}$
\end{center}
\caption{
Spectrogram of the $m=4$ Fourier component in a window spanning 
from a) time $T=350$ to 950 Myr and b) from 750 to 1305 Myr.  
Similar to Figure \ref{fig:spec2}.  
With the exception of the bar (with $\omega = 0.016$ at early times
and 0.013 Myr$^{-1}$ at later times)
the spiral patterns seen here are strongest between their inner and outer $m=4$ 
Lindblad resonances (as seen via intersections with green lines
showing $4 \omega \pm \kappa $.  
Higher frequency features are seen above
the bar's angular frequency at both early and later times.
\label{fig:spec4}
}
\end{figure*}

\begin{figure*} 
\begin{center}
$\begin{array}{cc}
\includegraphics[angle=0,width=3.5in]{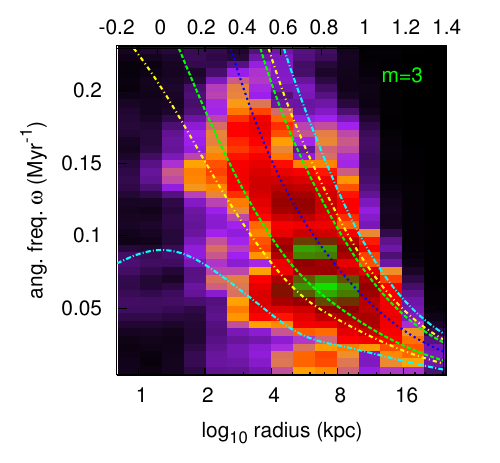} &
\includegraphics[angle=0,width=3.5in]{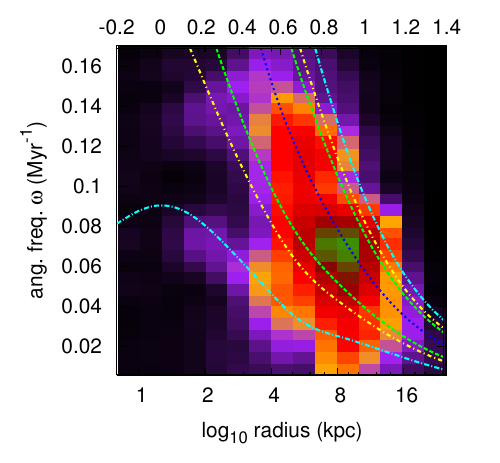}
\end{array}$
\end{center}
\caption{
Spectrogram of the $m=3$ Fourier components in a window spanning 
from a) time $T=350$ to 950 Myr and b) from 750 to 1305 Myr.  
Similar to Figure \ref{fig:spec2} but constructed using 
the $m=3$ Fourier components
instead of the $m=2$ ones.
Overplotted are seven lines that are three 
times the following angular frequencies
in order of bottom to top
$\Omega - \kappa/2$, $\Omega - \kappa/3$, $\Omega - \kappa/4$, 
$\Omega$, $\Omega + \kappa/4$, $\Omega + \kappa/3$, and $\Omega + \kappa/2$.
The dominant three-armed pattern
extends between its $m=3$ inner Lindblad resonance
and its $m=3$ outer Lindblad resonance at both early and later times.    
The bar itself on this plot has an angular frequency
of $\omega_b \approx 0.120$ at early times and 
$\omega_b \approx 0.105$~Myr$^{-1}$ at later times.  
There is a possible  coupling between the corotation resonance
of the bar and the $m=3$ inner Lindblad resonance 
of the three-armed spiral pattern as these two resonances are
close.
\label{fig:spec3}
}
\end{figure*}

\begin{figure*} 
\begin{center}
$\begin{array}{cc}
\includegraphics[angle=0,width=3.5in]{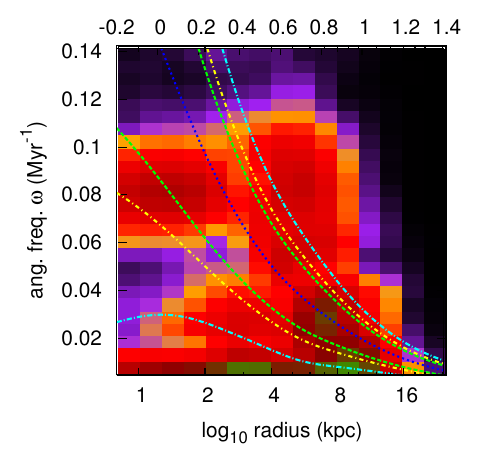} &
\includegraphics[angle=0,width=3.5in]{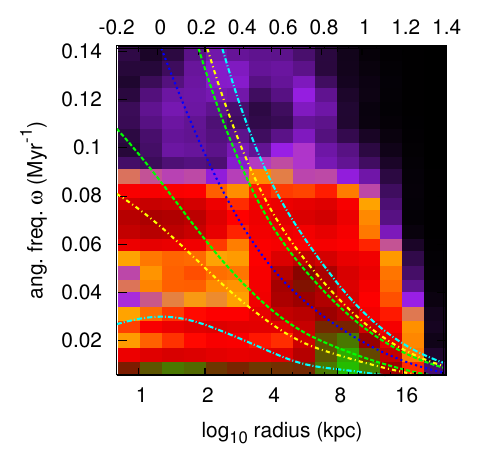}
\end{array}$
\end{center}
\caption{
Spectrogram of the $m=1$ Fourier component in a window spanning 
from a) time $T=350$ to 950 Myr and b) from 750 to 1305 Myr.  
Similar to Figure \ref{fig:spec3}.  
The bulge centroid position was not subtracted previous
to measurement of this spectrogram.
On the left the bright feature
at $\omega=0.08$ Myr$^{-1}$ is associated with the bar.  
The bar can be seen in the $m=1$
spectrogram because it moves with the lopsided oscillation.
Low frequency features may be symptomatic of non-linear coupling between waves.
\label{fig:spec1}
}
\end{figure*}

After $T\sim 400$ Myr the bar becomes stable and then evolves more 
slowly during the rest of the simulation. 
Well defined frequencies are seen in spectrograms after bar formation. 
Figures \ref{fig:spec2} shows spectrograms
constructed from $m=2$  Fourier components, whereas 
Figure \ref{fig:spec4}, \ref{fig:spec3} and \ref{fig:spec1} show spectrograms
constructed from $m=4$, 3 and 1 Fourier components, respectively.
The spectrogram from Figure \ref{fig:spec2}a is measured from 
the middle of the simulation
whereas \ref{fig:spec2}b is measured from the second half of the simulation. 
The same time periods are used for Figures \ref{fig:spec4} -- \ref{fig:spec1}.
We can see from a comparison of Figure \ref{fig:spec2}a and Figure 
\ref{fig:spec2}b (also see Figures \ref{fig:dens} and \ref{fig:l2})  that the bar
grows longer and its rotation rate slowly decreases during the simulation. 
At 400 Myr the bar is only 6 kpc long (semi-major axis) but by 700 Myr its semi-major
axis has increased to 11 kpc.

Pattern speeds of waves can be estimated from a spectrogram by dividing the
angular frequency of a strong feature by the integer $m$
used to make the spectrogram.
In our spectrograms we plot the angular frequency, 
$\omega$, rather than the pattern speed $\Omega=\omega/m$, 
following most previous works
(e.g, \citealt{sellwood88,masset97}).
Estimated pattern speeds for waves seen in the presented spectrograms
 are listed in Table \ref{tab:tab1}.
The strongest pattern seen in 
the $m=2$ spectrogram shown in Figure \ref{fig:spec2}a is the bar wave  
that has an angular frequency of $\omega_B \approx 0.080$ Myr$^{-1}$  
corresponding to a pattern speed of $\Omega_B \approx 0.040$ Myr$^{-1}$.
There are slower patterns at larger radii seen both in the middle 
and later parts
of the simulation.  These correspond to 
two-armed spiral patterns.   Figure \ref{fig:spec3} 
showing the $m=3$ spectrograms shows that there is also at least one strong
three-armed pattern present in the simulation.

Figure \ref{fig:spec2} and \ref{fig:spec4} 
show curves corresponding to the angular frequencies 
$m (\Omega \pm \kappa/2), m(\Omega \pm \kappa/4), m \Omega$ with $m=2$
and $m=4$, respectively where $m$ is that used to compute
the spectrogram.  Each pair of lines (when there is a $\pm$ in the
above listed frequencies) is the same color in the Figures.
These lines can be used to identify inner and outer Lindblad Resonances.
Figure \ref{fig:spec3} and \ref{fig:spec1} in addition show curves 
corresponding to the angular frequencies 
$m(\Omega \pm \kappa/3)$, with $m=3$ and $m=1$, respectively.
Here $\Omega(r)$ and $\kappa(r)$ are the angular rotation rate
and the epicyclic frequency.
We computed these frequencies using a rotation curve 
generated from the azimuthally averaged midplane gravitational potential 
measured from the positions of all the massive particles at time $T=500$ Myr.   

A Lindblad resonance (LR) with a pattern with speed, $\Omega_p$,
and integer $m$ occurs where 
\begin{equation} 
m (\Omega -\Omega_p)  = \pm \kappa
\end{equation}
\citep{lindblad26}.
The sign in the above determines whether the resonance is
an inner or outer Lindblad resonance (ILR or OLR).
A corotation resonance (CR) occurs where $\Omega = \Omega_p$,
 or the angular rotation rate equals the pattern speed.
Here we refer to each resonance by its integer, here denoted $m$.  
Other works refer to the $m=4$ ILR as the 4:1 ILR (e.g., \citealt{meidt09})
or the 4/1 ILR (e.g., \citealt{rau99}).  We also use the notation
ILR$_4$ to refer to the $m=4$ ILR.
The radii of the estimated Lindblad and corotation resonances for 
pattern speeds estimated from the spectrograms 
are also shown in Table \ref{tab:tab1}.
Resonance locations are estimated using the rotation curve computed 
from the azimuthally averaged density distribution
at time $T=500$ Myr.
The bar pattern also contains non-zero Fourier components in its density
distribution with $m\ne 2$.
As the bar lacks symmetry when rotated $180^\circ$, 
or when reflected about its major or minor axis,
it also contains non-zero odd-$m$ Fourier components
in its density distribution.

We see in Figure \ref{fig:spec2}a that the bar pattern
extends out to 
 its corotation radius where
the bar pattern intersects $2 \Omega_B$ (the dark blue line) at 
about 6 kpc. 
The pattern is seen past the bar's corotation radius in the spectrogram,
consistent with previous studies (e.g., see Figure 3 by \citealt{rau99}).
Between the corotation resonance and the bars $m=2$ outer Lindblad resonance
(OLR$_2$), periodic orbits can be elongated and oriented perpendicular to the bar.

Recent studies of the Milky Way have revealed two bar like structures,
a long thin bar \citep{benjamin05} at about $45^\circ$ from the Sun/Galactic center line 
and half length 4.4 kpc and a shorter
or traditional thicker bar (or triaxial bulge component) at about $15^\circ$ \citep{vanhollebeke09}.
Estimates of the bar's pattern speed are in the range 50-55 km s$^{-1}$ kpc$^{-1}$ or $\sim 0.055$Myr$^{-1}$
\citep{dehnen00,minchev07,gardner10}.  As shown by \citet{gardner10} the Hercules stream is consistent
with either long or short bar as long as the pattern speed places the solar neighborhood just outside the bar's OLR$_2$.
Our simulated bar has with half length $\sim 5$ kpc and so
is longer than the Milky Way's long bar but this is consistent with
its slower pattern speed at 0.040 Myr$^{-1}$ or 40 km s$^{-1}$ kpc$^{-1}$.
In our simulation a position corresponding to that of the solar neighborhood
 would be at a galactocentric radius of about 10 kpc, just outside the bar's   OLR.  

At both earlier and later times, features at lower pattern speeds are also
seen in the $m=2$ spectrograms.
A peak just past the end of the bar is seen
with a pattern speed of $\Omega_s \sim 0.030$~Myr$^{-1}$ 
at earlier times
(Figure \ref{fig:spec2}a)  and at $\Omega_s  \sim 0.022$ at later times
(Figure \ref{fig:spec2}b).
This would correspond to an inner two-armed spiral pattern. 
This pattern appears to end within its own corotation radius at 7 or 9 kpc (early and late, respectively).
There is also a slower  outer two-armed spiral pattern primarily seen at 
early times with a pattern speed 
$\Omega_s \sim 0.015$ (though possibly also at later
times with $\Omega_s \sim 0.010$). 
This pattern also ends near its own corotation resonance.

There is a weak and narrow higher frequency peak in the $m=2$ spectrograms
near the end of the bar
at an angular rotation frequency of $\omega \sim 0.13$ 
and $\omega \sim 0.11$ Myr$^{-1}$ at
early and late times, respectively.
Unlike the slower frequency in the $m=2$ spectrograms that
end near their corotation radii, this one extends between
its corotation radius and its outer Lindblad resonance.
There is a similar higher frequency feature in the $m=4$ spectrograms
(Fig. \ref{fig:spec4}) at $\omega \sim 0.23$ and 0.17 Myr$^{-1}$ at
early and later times, respectively.

The $m=4$ spectrograms shown in Figure \ref{fig:spec4} look remarkably 
similar to the $m=2$ spectrograms (Figure \ref{fig:spec2}). 
Two armed and oval density distributions often contain $m=4$ components so 
this is not necessarily surprising.  The higher frequency features
near the end of the bar are also seen in the $m=4$ spectrograms.
With the exception of the bar itself, the 
peaks in the $m=4$ spectrogram extend between their inner and
outer $m=4$ Lindblad resonances (shown as green lines).
This differs from the two-armed spiral patterns that end within
 their corotation radii.
In the $m=4$ spectrogram the inner spiral pattern appears split into
two different peaks.   
We may be able to better measure angular
frequencies in the $m=4$ spectrogram than the $m=2$ spectrogram because
the angular frequencies are higher and so less affected by the precision 
limit set from the width of the time window used to compute the spectrograms. 

Unlike the inner and outer two-armed patterns, the three-armed pattern both
at early and later times extends
through its corotation resonance.  This pattern
(see yellow lines in the $m=3$ spectrograms shown in Figure \ref{fig:spec3}) 
extend between its $m=3$ inner and outer Lindblad resonances. 
The pattern speed is difficult to measure at early times because
there are two nearby features in the spectrogram.  Both at 
early and later times the pattern speed of the three-armed
spiral is about $\Omega_s \sim 0.023$~Myr$^{-1}$.

As noted by \citet{shaviv03} (see their Table 3) pattern speeds
estimated for spiral patterns in the solar neighborhood
fall into two groups.   One group has pattern speed $\sim 27$ km s$^{-1}$ kpc$^{-1}$ similar to the
angular rotation rate of a particle in a circular orbit, $\Omega_\odot$, at the solar galactocentric
radius, $R_\odot$, and the other group has a pattern speed
$\sim 14$ km s$^{-1}$ kpc$^{-1}$.    
The mock solar neighborhood position in our simulation lies outside the inner
two armed spiral pattern at 0.030 Myr$^{-1}$, but is affected by
the three armed spiral wave with pattern speed 0.023 Myr$^{-1}$ or $\sim 23$ km s$^{-1}$ kpc$^{-1}$
and the outer two armed spiral with a pattern speed of 0.015 Myr$^{-1}$ or $\sim 15$ km s$^{-1}$ kpc$^{-1}$.
If the spiral waves present in the solar neighborhood are driven by the Galactic
bar then it might be interesting to match the faster pattern speed measured
in the solar neighborhood with 
a three armed pattern driven by the bar and the slower pattern speed with a two armed
pattern also possibly linked to the bar.  However,
with the exception of \citet{naoz07}, few studies have considered interpretation of
the solar neighborhood in terms of two or more patterns.

Previous works have seen spiral structure in N-body or particle
mesh simulations at pattern speeds that differ from the bar's 
(e.g., \citealt{sellwood88,patsis99,rau99,voglis06}).
\citet{masset97} interpreted this behavior 
in terms of non-linear mode coupling \citep{tagger87,sygnet88},
whereas \citet{sellwood89} interpret the differences
in pattern speeds in terms of transient behavior and a groove instability.
The spiral patterns in the simulations by \citet{sellwood88} extend
from their $m=2$ inner Lindblad resonances to
their outer Lindblad resonances (see their Figure 1).
However, the patterns of the two spiral density waves seen by \citet{patsis99}
in a simulation lacking a bar
extend from their $m=2$ inner to 
their outer $m=4$ Lindblad resonances.
Both of these situations differ from the two-armed structures
in our simulations that end within their corotation radii.
The simulations by \citet{rau99} are closest in behavior
to ours, showing multiple spiral patterns
extending past the bar, with some patterns ending near
or within their corotation radii.  

As the spiral and bar patterns in our simulations are moderately long lived,
the spiral patterns could be driven by the bar
\citep{yuan97,masset97,rau99}. 
\citet{rau99} accounted for the ranges and speeds
of spiral patterns using the non-linear wave coupling scenario.
They found that they could often associate
the beginning of a slower outer pattern with the location
of one of its resonances 
that is also a resonance with an inner faster bar or other spiral pattern.
The overlap between the resonances relates one pattern speed to
the other and couples one perturbation to the other.
\citet{rau99} found examples of CR/ILR$_4$,
CR/ILR$_2$, OLR$_2$/CR and  OLR$_2$/ILR$_2$
couplings between spiral and bar waves and between 
two different spiral density waves.

For our system there is a possible connection between
the bar's CR and the $m=3$ ILR of the three-armed spiral armed
wave.  A comparison of resonance
locations (see Table \ref{tab:tab1}) shows
that the bar's corotation resonance lies close 
to the three-armed spiral's $m=3$ ILR.  
The radial distance between these resonances is small compared
to the error from measurement in the spectrogram.
There is also a possible coupling between
other resonances.
It is difficult to differentiate
between possibilities by resonance locations alone
as driving could occur in a spatially
broad region near a resonance.  As suggested by \citet{masset97}
we can test for mode or wave coupling by searching for waves at beat frequencies
though  \citet{rau99} found that expected beat waves 
are not always detectable.

\subsection{Non-linear Coupling of Waves}

The scenario introduced by \citet{tagger87} suggests that there are 
strong non-linearities in the stellar response when resonances coincide.
Two waves (e.g., bar and spiral) couple to a third one at a beat frequency.
For example we consider azimuthal wavenumber $m_B$  associated
with a bar perturbation at pattern speed $\Omega_B$ and 
so frequency $\omega_B = m_B \Omega_B$.
We consider azimuthal wavenumber $m_S$ associated with a spiral pattern
at pattern speed $\Omega_s$ and so angular frequency $\omega_S = m_S \Omega_S$.
A third wave can be excited with an 
azimuthal wavenumber $m = m_B + m_S$ and 
angular frequency $\omega = \omega_B + \omega_S$ or with 
wavenumber $m = m_B - m_S$ and $\omega = \omega_B - \omega_S$.

An example explored by \citet{tagger87} is a bar and spiral mode, 
both with $m=2$, coupled to a slow axisymmetric mode with $m=0$
and a faster four-armed one with $m=4$ \citep{masset97}.
Like \citet{masset97} we see  features in the $m=4$ spectrograms at frequencies
above that of the bar, and these can be considered predictions of
the wave-coupling scenario.  Unfortunately  the number of
features present in the $m=2$ and $m=4$ spectrograms is 
high and the imprecision of measured
 frequencies makes it difficult
to definitively determine whether there is coupling between $m=2,4$ waves and a slow axisymmetric mode.

As our simulation contains a bar ($m=2$), a lopsided $m=1$ motion
and a three armed wave we can
 consider coupling between these three waves.
To create the $m=2$, 3 and 4 spectrogram 
we first subtracted the bulge centroid,  
however the galaxy center moves during the
simulation.  In Figure \ref{fig:spec1}, showing the $m=1$ spectrograms
but constructed in the inertial frame,
 we see features associated with the major patterns 
identified in the other spectrograms.
For example we see a peak at $\omega = 0.08$ Myr$^{-1}$
that coincides with the bar's angular frequency.  One interesting 
peak is at an angular rotation rate
of $\omega \sim 0.010$ that is also seen in the $m=3$ spectrogram at a radius of about 8 kpc 
($log_{10} r = 0.9$).  At earlier times there is power at $\omega\sim 0.03$ 
in the $m=1$ spectrogram
that is broader than the similar feature
seen in the $m=2$ spectrogram.   This suggests
that there is sufficient power at low frequencies in the lopsided
wave that it could couple to other waves.

Since there is power at low frequencies in lopsided motion,
we can consider the possibility of 
a coupling between a slow lopsided mode and the bar. Here
the index and pattern speed of the lopsided wave is
 $m_L=1, \Omega_L$ and that of the bar is $m_B=2, \Omega_B$.
The sum leads to a three-armed spiral pattern with angular frequency 
$\omega_S = \Omega_L + 2 \Omega_B$
or pattern speed $\Omega_S = {1\over 3}\left(\Omega_L + 2 \Omega_B\right)$.
If we assume that this corresponds to the three-armed pattern with 
$\Omega_S \sim 0.028$ then we can solve for the angular frequency of 
the lopsided mode $\Omega_L \sim 0.01$.  There may be power at this 
frequency as there are low frequency features in the $m=1$ spectrogram.
Thus, a lopsided slow mode could be coupling the bar and the three 
armed spiral waves.


A variety of resonant overlaps or non-linear wave
couplings are possible and may account for the driving of the patterns
present in the simulation.  
If we could more precisely measure frequencies of the patterns 
then we would better determine which associations are most 
important or likely.   
Unfortunately peaks in the spectrograms are sensitive to the range 
(in time) of simulation data used to construct the spectrogram 
(e.g., compare Figure 
\ref{fig:spec2}a to Figure \ref{fig:spec2}b).  
Longer time series are required to make more accurate frequency 
measurements.  Since the pattern speeds are not fixed,  
increasing the size of the time window does not increase the precision 
of a frequency measurement.  This is particularly the case for slow patterns 
that might be important for the non-linear coupling models.

In summary, we find that the spectrograms are rich in small features.
The shape of the spectrograms seen at early times resembles that at
later times implying that the frequencies of these smaller features 
are slowing along with the bar pattern.  
Even though there are many waves present, they are not stochastically
appearing and disappearing but slowly evolving.
Lindblad resonances associated with one wave are likely
to be coincident with resonances of another wave present in the simulation.
Together these findings suggest that the waves present
in the simulation are coupled due to non-linear interactions.
The most convincing and strongest coupling 
is that between the bar, the lopsided wave and the three-armed
spiral wave.   The multitude of small features present
in the spectrograms suggest that additional couplings
are present.  The presence of features in the spectrograms at frequencies above
that of the bar is consistent with the wave-coupling scenario.

\section{Structure in the $uv$ Plane}

We first describe how we construct velocity distributions in local 
neighborhoods.  We then discuss possible relations between
features seen in these velocity distribution and
the spiral or bar patterns.

\subsection{Velocity $uv$ distribution histograms 
in local neighborhoods}

We construct velocity distribution histograms in different local neighborhoods 
in the following manner.  Consider a neighborhood centered about a
point in the disk plane with coordinates
$(x_0, y_0, 0)$ or in cylindrical coordinates $(r_0, \phi_0, 0)$.
We will refer to neighborhoods by the cylindrical coordinates 
of their centers; ($r_0,\phi_0$).
We consider a star or simulated particle in the neighborhood of this center 
point if its coordinates $(x, y, z)$ 
have distance in the plane, $R = \sqrt{(x-x_0)^2 + (y-y_0)^2} < r_0 f$
where $f$ is a dimensionless 
scaling factor setting the size of the neighborhood.
We compute the radial and tangential components of the velocity
of each star with respect to the vector to the galactic center
defined by the central coordinates of the neighborhood, $(x_0,y_0,0)$.  
The $u$ component is $-1$ times the radial velocity component
in this galactic coordinate system. The $v$ component
is the tangential component in this coordinate system minus the circular
velocity estimated from the azimuthally averaged midplane rotation curve
computed at time $T=500$ Myr.

Histograms showing the $uv$ velocity distribution 
are computed from the disk particles in each neighborhood.
The computed histograms are effectively 
coarse-grained phase space distributions.
The larger the neighborhood, the larger number of stars in the neighborhood. 
If there are too few stars in the neighborhood then the velocity distribution
is noisy. If the neighborhood is large then structure present 
on small spatial scales may be masked or smoothed by the coarse graining.  
An extreme example would be if the neighborhood were so large that it contained
two spiral arm density peaks.  

A series of velocity distributions ($uv$ histograms) is shown in Figure
\ref{fig:page120} at simulation time $T=$600~Myr. 
Velocity distributions are extracted from neighborhoods centered
at 8 equidistant angles and at 6 radii spaced logarithmically.
Each row contains histograms from neighborhoods at a particular galactocentric
radius.  The  lowest row has $r_0=4.1$ and the
top row $r_0=12.5$.  Each radius differs from the previous
by a factor of 1.25.
Each column shows neighborhoods with a particular galactocentric angle with
the leftmost column aligned with the bar.  Each neighborhood is separated
in angle by 45$^\circ$ with angle increasing counter-clockwise in the
galaxy plane.
Each neighborhood has a size set by the factor $f$ setting the maximum
distance of stars included in the neighborhood from its
center.   For the neighborhoods shown in Figure \ref{fig:page120} the ratio
of the neighborhood radius to $r_0$ is $f= 0.1$.
Positions and sizes of each neighborhood
are shown in Figure \ref{fig:dcirc120} in both Cartesian and 
polar coordinates.
At $r_0 = 8$ kpc the radius of the neighborhood corresponds to
0.8 kpc which is small compared to some N-body studies 
(e.g., \citealt{gomez10} who used a neighborhood radius of 2.5~kpc) but
larger than the distance of stars in solar neighborhood samples
(e.g., \citealt{siebert11}).

We have compared the velocity distributions created with 
$f=0.1$ and with $f=0.05$ with smaller neighborhoods.  
A comparison at time 800 Myr and at a radius of 10 kpc is shown 
in Figure \ref{fig:comp}.  We also show a velocity distribution
constructed of massive disk particles only to illustrate that we
see no evidence for significant differences between massive and massless
disk particle populations.
The velocity distributions in smaller neighborhoods
are noisier as would be expected since each neighborhood 
contains fewer particles.
The clumps in the plots are slightly sharper, otherwise 
the morphology is almost identical. The density distribution 
is relatively smooth (see Figure \ref{fig:dcirc120}) 
implying that the velocity dispersion is not low in the disk
(though it is lower at larger radii than at smaller radii). 
A simulation of a colder disk with more tightly
wound spiral arms might show features
in the velocity distributions that are more strongly
dependent on the size of the neighborhood.

In Figure \ref{fig:page120}
we see that multiple clumps exist in the local velocity distribution
at all radii in the galaxy. 
Velocity clumps were previously seen in a local neighborhood
of the N-body simulation presented by \citet{fux00}.
Since our simulation lacks mergers,
the clumps in the velocity distribution
are caused purely by dynamical phenomena in the disk (as also concluded
by \citealt{fux00}).   
Clumps in the velocity distribution in one neighborhood
are often also present in the nearby neighborhood at a larger or smaller radius
but at a shifted $v$ velocity.  For example at an angle of 315$^\circ$
there is an arc with a mean velocity of $v\sim 50$~km/s for 
neighborhood radius $r_0=4.1$
that has a lower velocity of $v \sim 10$~km/s for at $r_0=5.1$.
We can consider the possibility that a single halo particle could be responsible for one of
these features.   The separation between the two neighborhoods mentioned above is a kpc.
However the circular velocity at a distance of a kpc from one of our halo particles 
is only  2 km/s.  Thus a single halo particle passing through the disk cannot account for the correlated
and broad structures seen in the velocity distributions.
We will discuss the relationship between $v$ 
and the mean orbital or guiding radius below when we consider
the relation between velocity components and orbital properties. 

\subsection{Velocity distributions in arm and interarm regions}

\begin{figure*}  
\includegraphics[angle=90,width=6.0in]{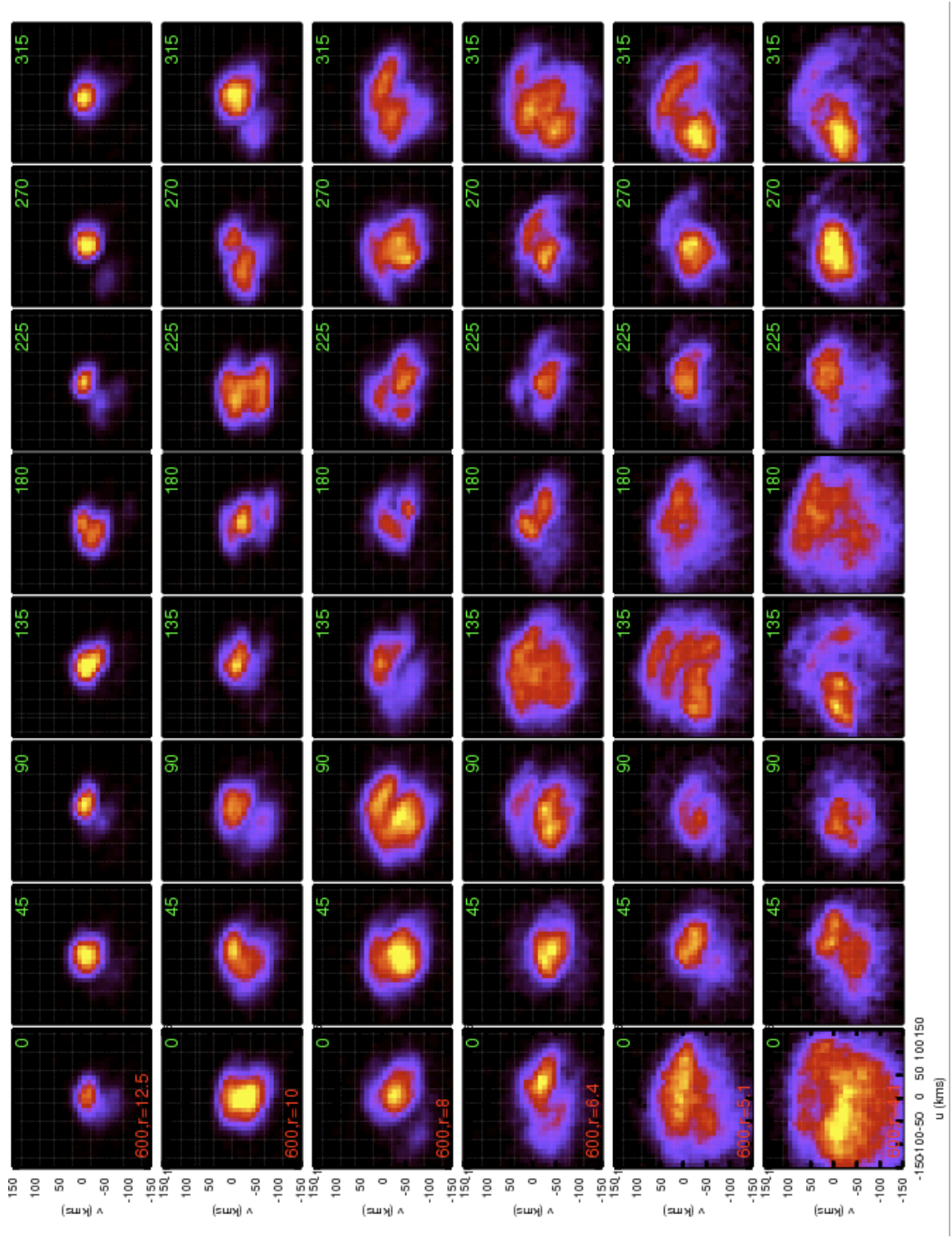}
\caption{
Local velocity distributions at different positions in the outer galaxy
at a simulation time $T=600$ Myr. 
Velocity distributions are extracted from neighborhoods centered
at 8 equidistant angles and at 6 radii spaced logarithmically.
Each panel shows the distribution in $u$ (x-axis) versus $v$ (y-axis).
Each row shows neighborhood at a particular galactocentric 
radius with the lowest row with $r_0=4.1$ kpc and the 
top row with $r_0=12.5$ kpc.  
Each column shows neighborhoods with a particular galactocentric angle with 
the leftmost aligned with the bar.  Each neighborhood is separated
in angle by 45$^\circ$ with angle increasing counter-clockwise in the
galaxy plane.  
Positions and sizes of each neighborhood
are shown in Figure \ref{fig:dcirc120}.
Here we see that clumps exist in the velocity distribution
at all radii in the galaxy.  Features in the velocity
distribution at one neighborhood 
are often seen in the nearby neighborhood at a larger or smaller radius
but at a shifted $v$ velocity.  
\label{fig:page120}
}
\end{figure*}

\begin{figure*} 
\begin{center}
$\begin{array}{cc}
\includegraphics[angle=0,width=3.5in]{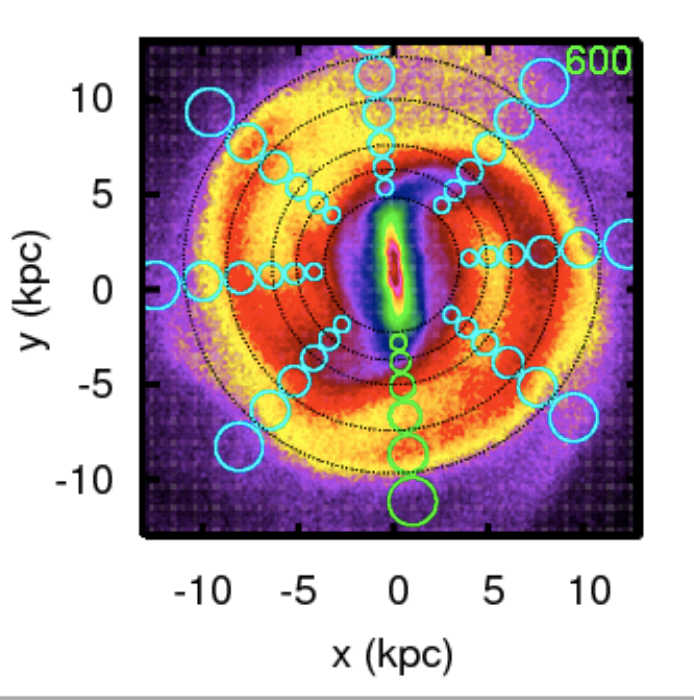} &
\includegraphics[angle=0,width=3.5in]{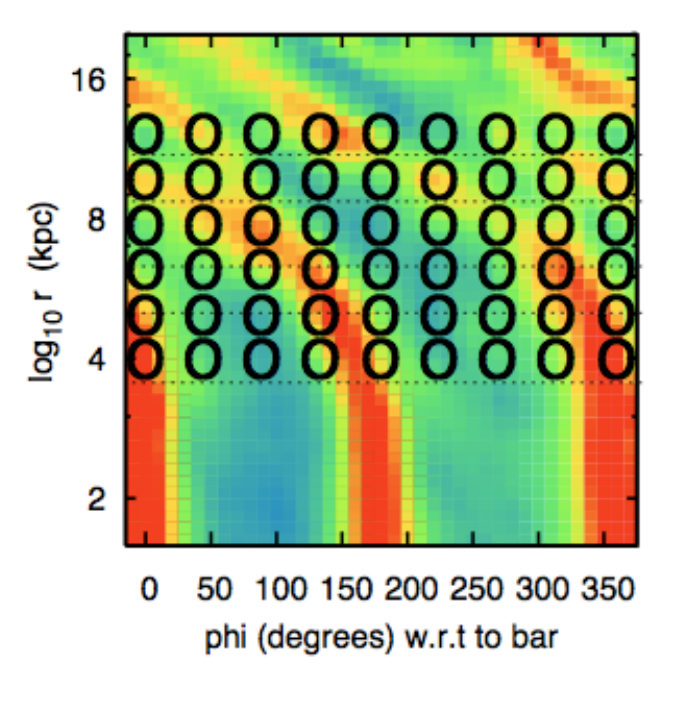} 
\end{array}$
\end{center}
\caption{
The positions of the neighborhoods used to construct velocity distribution
histograms shown in Figure \ref{fig:page120}.
a) The neighborhoods are shown in a Cartesian coordinate system.
The green circles show the angle of the neighborhoods 
oriented along the bar and referred to as   
angle zero (with respect to the bar) and
are shown in the leftmost column of Figure \ref{fig:page120}.
In Figure \ref{fig:page120} the angles of each column, from left to right,
correspond to positions with angle increasing by 45$^\circ$ 
in the anti-clockwise direction on this plot. 
Radii of changes in spiral arm pitch angle or discontinuities
in spiral structure estimated from the polar plots are shown
as dotted black circles.
b) The approximate locations of the neighborhoods shown on a projection 
of angle versus logarithmic radius (the differential density distribution is
constructed as in Figure \ref{fig:l2}).  
The positions of the neighborhoods on this plot are in the same order
as the velocity distribution panels shown in Figure \ref{fig:page120}.
Black dotted horizontal lines are shown at radii where there are 
changes in spiral arm pitch angle or there are discontinuities 
in the spiral arms.
Many of these lines  correspond to resonances with patterns seen
in the spectrogram.  
\label{fig:dcirc120}
}
\end{figure*}

When looking at the velocity distribution in different
neighborhoods it is helpful to know which neighborhoods have a high surface density.
Figure \ref{fig:dcirc120}a and b also show the locations of neighborhoods 
 on the projected disk surface density used to construct the velocity
 histograms shown in Figure \ref{fig:page120}.
In Figure \ref{fig:dcirc120}b, projected in polar coordinates, 
the $x$ axis gives the azimuthal angle with respect to the bar.  
Thus each neighborhood shown
on this plot directly corresponds to the position of the 
$uv$ histogram in Figure \ref{fig:page120}. This makes it possible to see how 
the velocity distribution varies with position and between arm
and interarm regions. 

A comparison between Figure \ref{fig:page120} and \ref{fig:dcirc120}b 
allows us to see if the velocity distribution in 
high density regions differs from that at low density regions.
We find that when the density increases 
the velocity dispersion is also likely to increase.
The velocity distribution is wider when the neighborhood lies
on an arm or bar peak than in interarm regions.
Phase space density should be conserved
as particles orbit the galaxy, so this trend is expected. 

We can compare the velocity distribution on the leading (or pre-shock) side
of a spiral arm to that on the trailing (or post-shock) side.    
Our simulated galaxy rotates counter-clockwise or with increasing $\phi$. 
We denote the leading side of a spiral arm as that located at larger $\phi$
than the peak, or to the right in Figure \ref{fig:dcirc120}b 
(and counter clockwise
in angle from an arm peak in Figure \ref{fig:dcirc120}a). 
Observed galaxies tend to have dust lanes on the trailing side (as denoted here) of a spiral 
arm.  This happens because the spiral pattern is usually faster than local galactic rotation 
(in other words the corotation radius lies interior to the pattern).
In a frame moving with the gas on Figure \ref{fig:dcirc120}b, the pattern moves to the right and so shocks
would form on the left-hand or trailing side of each density peak. 

We see that arcs in the velocity distribution
with positive $u$ and $v$ are often seen on spiral arm
density peaks and on the leading (or pre-shock) side of an arm. 
These arcs are oriented with negative slope 
such that $v$ increases with decreasing
$u$.  Moving from the leading to the trailing side at a particular radius
(comparing histograms from right to left in Figure \ref{fig:page120})
the high $v$ arcs move from positive to negative $u$.
See for example $r=8$ and $\phi=90, 135^\circ$  or
$r=6.4$ and $\phi=315,0^\circ$  in Figure \ref{fig:page120}.
Interarm regions tend to lack arcs that are 
centered at high and positive $u,v$. 

In interarm regions the velocity dispersion 
is lower and there is sometimes a clump, separated from the dominant peak,
that is at negative $u$ and $v$.  The clump can be elongated in $u$,
resembling the Hercules stream in the solar neighborhood's
velocity distribution (e.g., \citealt{dehnen98,arifyanto06,gardner10,bovy10}).

\subsection{Gaps in the $uv$ histograms}

Figure \ref{fig:pages6} 
shows the neighborhoods at a radius of $r_0=$ 6.4 kpc 
for the same angles shown in 
Figure \ref{fig:page120} but for snapshots separated
by 50 Myr between 500 and 950 Myrs.
The angles here are given with respect to the bar and so
in a coordinate system rotating with the bar.  
At each angle with respect to the bar, clumps do not remain 
fixed as the system evolves. 
This implies that spiral structure influences the locations of gaps and
clumps in the velocity distribution.  
The gaps cannot be solely due to bar perturbations.

A stream similar to the Hercules stream at $v\sim -50$ km/s is present
in the velocity distribution at early times 
at angle 0 (aligned with the bar) at $r_0 = 6.4$ kpc.  Like the Hercules
stream (e.g., \citealt{dehnen99}) this arc or clump has a wide
$u$ distribution, a narrow $v$ distribution 
and is centered at a negative $u$ value.
However the bar's $m=2$ OLR (at about 9 kpc)
lies outside these neighborhoods, so this clump cannot be
associated with this resonance, 
Similar feature low $u,v$ clumps in the velocity distribution 
are also seen at other radii such as $r_0 = 8$ and 10 kpc 
but appearing at different times in the simulation and at different angles
with respect to the bar, consequently the
bar and resonances associated with it
cannot provide the only explanation for these features.

The spiral density waves have slower patterns than the bar, hence
density peaks associated with them move to the left (or to a smaller $\phi$)
as a function of time or moving downward on Figure \ref{fig:pages6}.
At one position as a wave peak passes through the neighborhood first
a velocity distribution typical of a leading side of an arm is seen
then one typical of a trailing side.
The velocity distribution contains a positive $u$ and $v$ feature
when on the leading side of a spiral feature.    The feature is 
centered at progressively
lower $u$ values as the arm passes by and the neighborhood 
moves to the trailing side. See for example the 
right most column and at an angle of 315$^\circ$ 
between times 500 to 700 Myrs (top 5 rows).

While clumps come and go in time, we find that 
the gaps between the clumps tend to remain
at the same $v$ values.  For example
consider the right most column of Figure \ref{fig:pages6}.  Two
gaps are seen with different degrees of prominence at differ times.
We recall from Figure \ref{fig:page120} that gaps at one neighborhood
radius were often seen at another neighborhood radius but shifted in $v$.
To compare a gap seen in a neighborhood at one radius to that
seen in a neighborhood at another radius we can consider
motions of particle orbits.

\section{Interpreting structure in the $uv$ plane}

To relate structure in local velocity distributions it is helpful
to recall models for orbits.  We first consider epicyclic motion
in the absence of perturbations, then review the orbital dynamics to first order
in the perturbation strength in the presence of perturbations.
In both case we consider the relation between $uv$ velocities
in a local neighborhood and quantities describing the orbit
such as a mean or guiding radius and an epicyclic amplitude.

In the absence of perturbations from
spiral arms, the motion of stars in the disk of a
galaxy can be described in terms of radial or epicyclic oscillations
about a circular orbit \citep{lindblad26,kalnajs79}.  
It is useful to specify the relation between the observed
velocity components $u,v$ and the parameters describing
the epicyclic motion (e.g, \citealt{fux00,fux01,quillen05})  or 
the mean radius or guiding radius, $r_g$, and the epicyclic amplitude.
The energy of an orbit in the plane of an axisymmetric
system (neglecting perturbations from spiral structure) with a  flat
rotation curve is
\begin{equation}
E(u,v) = {(v_c + v)^2\over 2}  + {u^2 \over 2} + v_c^2 \ln{r} + {\rm constant}
\label{uv}
\end{equation}
where the potential energy, $\ln{r}$, is that appropriate for a flat rotation
curve, $r$ is the Galactocentric radius and $v_c$ is the circular velocity.

With an epicyclic approximation  we can write the energy
\begin{equation}
E = v_c^2 \left[{1\over 2} + \ln{r_g}\right] + E_{epi}
\label{epi1}
\end{equation}
where the term on the left 
is the energy of a star in a circular orbit
about a guiding radius $r_g$ and $E_{epi}$ is the energy from
the epicyclic motion,
\begin{equation}
E_{epi} = {u^2 \over 2} + {\kappa^2 (r-r_g)^2 \over 2}
= {\kappa^2 a^2 \over 2}.
\label{epi2}
\end{equation}
Here $a$ is the epicyclic amplitude and $\kappa$ is the epicyclic
frequency at the guiding radius $r_g$.

We now consider stars specifically in a neighborhood of galactocentric radius
$r_0$, restricting us to a specific location in the galaxy.
Setting the energy
equal to that written in terms of the epicyclic motion
using equations (\ref{epi1}, \ref{epi2}),
we solve for the guiding radius, $r_g$.   
It is convenient to define the distance between
the guiding or mean radius and the neighborhood's galactocentric radius, 
$s=r_g -r_0 $.
To first order in $v$ and $s$ we find that a star in the neighborhood 
with velocity components
$u,v$ has a guiding radius with
\begin{equation}
{s\over r_0} \approx  {v \over v_c}
\label{eqn:s}
\end{equation}
and epicyclic amplitude $a$
\begin{equation}
{a\over r_0} \approx v_c^{-1} \sqrt{{u^2\over 2} + {v^2}}.
\label{eqn:aepi}
\end{equation}
The factor of a half is from $\kappa^2/\Omega^2$ equivalent to 2 when 
the rotation curve is flat.
An angle describing the phase in the epicycle is 
\begin{equation}
\phi 
          \sim {\rm atan} \left({ -u \over \sqrt{2} v} \right)
\label{eqn:theta1}
\end{equation}
(consistent with equations 2-4, \citealt{quillen03}, for a flat rotation curve).
The epicyclic phase angle $\phi =0$ at apocenter.

The above three relations allow us to relate the velocity components
$u,v$ for stars in the solar neighborhood,
to quantities used to describe the epicyclic motion; the guiding
radius, epicyclic amplitude and phase angle. 
Particles with positive $v$ have $s>0$
and so guiding radii that are larger than the radius of 
the neighborhood, $r_0$. 
Particles with negative $v$ have $s<0$
and so are expected to spend most of their orbits inside $r_0$.  The
distance from the origin, $u=v=0$, determines the epicyclic amplitude.

It may also be useful to write
\begin{equation}
{v \over v_c} =  {r_g \over r_0} -1,
\label{eqn:vfind}
\end{equation}
where we have used our definition for $s$ in equation \ref{eqn:s}.
This equation shows that to first order $v$ sets the guiding radius of a
particle.  This is useful because the guiding radius sets the
approximate angular rotation rate and rotation period of the orbit.
The guiding radius also sets the approximate epicyclic frequency (or period)
of the particle's orbit.  Resonances occur where sums of integer multiples
of these periods are equal to an integer multiple of a pattern speed.
Thus the $v$ value of clumps or gaps in the velocity 
distribution may also allow us to locate resonances.

\subsection{Spiral Perturbations to First Order in the $uv$ plane}

The above approximation neglects the effect of bar or spiral perturbations.
Spiral  waves give a time dependent perturbation to the gravitational potential 
that can be approximated by a single Fourier component depending on  a single frequency.
 For a logarithmic spiral perturbation with $m$ arms we can consider
 $V(r,\theta,t) = V_m \cos(m (\theta - \Omega_pt  + \alpha \ln r))$
where the  pitch angle  $\alpha = -d\theta/dr$.
To first order in the perturbation strength $V_m$ the orbit of a star in the galactic
midplane with mean or guiding radius $r_g$
\begin{eqnarray}
r(r_g,\theta, t) &=& r_g + a \cos (\kappa_g t + \phi_0) + \nonumber \\
&& C(r_g) \cos (m(\theta - \Omega_p t + \alpha \ln r_g + c_0))
\label{eqn:rrg}
\end{eqnarray}
(\citealt{B+T}; equation 3-119a)
where $a$ is an epicyclic amplitude, and $c_0, \phi_0$ are  phase angles.
The forced epicyclic amplitude in the WKB approximation $|C(r_g)| \approx \left|{\alpha V_m \over r_g \Delta_g}\right|$
 where
 $\Delta_g = \kappa_g^2 - m^2(\Omega_g - \Omega_p)^2$ represents the distance to a Lindblad resonance
and is evaluated at the guiding or mean radius, $r_g$ and $\kappa_g$ and $\Omega_g$ are the epicyclic 
frequency and angular rotation rate at $r_g$.
The denominator $\Delta_g$ becomes small near the Lindblad resonance and the first order
approximation breaks down.  A higher order approximation can be used
to show that closed orbits exist on both sides of resonance and that epicyclic
amplitudes do not become infinite on resonance (e.g., \citealt{cont75}).
   
The orbital motion (equation \ref{eqn:rrg}) consists of an epicyclic motion with amplitude $a$ and a phase
$\phi_0$ and a forced motion aligned with the potential perturbation and moving with it.
A population of stars would have a distribution in both $a$ and $\phi_0$.
When $a=0$ and $m=2$ the orbits are ellipses.  The orbit with $a=0$ is periodic or closed
in the frame rotating with the spiral pattern.
The mean value,  $\bar a$, sets the velocity dispersion in the disk.  When $\bar a < |C|$
 the orbits are nearly closed in the frame rotating with the potential perturbation and we can 
 consider the disk to be cold.

The radial and tangential velocity components can be computed
\begin{eqnarray}
v_r(r_g,\theta,t) &=& - a \kappa_g \sin (\kappa_g t + \phi_0) 
\label{eqn:urg} \\ 
&& - C(r_g) m (\Omega_g - \Omega_p) \sin (m(\theta - \Omega_p t + \alpha \ln r_g + c_0)) \nonumber
\\
v_\theta(r_g,\theta,t)  &=& r_g \Omega_g - a \Omega_g \cos (\kappa t + \phi_0) \label{eqn:vrg}  \\
&& -C (r_g)\Omega_g   \cos (m(\theta - \Omega_p t + \alpha \ln r_g +c_0)).\nonumber
\end{eqnarray}
(from equation 3.117 by \citealt{B+T}).
When the rotation curve is flat $r_g \Omega_g$ is independent of radius and the $uv$ velocity
components  $u = -v_r$ and $v = v_\theta - r_g \Omega_g$.

We note that the equations for $r(r_g,\theta,t)$ and $v_\theta(r_g,\theta,t)$ contain similar terms but
with the opposite sign.    
A shift in $v$ corresponds to a shift in radius and the mean orbital
radius can be roughly estimated from the $v$ component as was true in the case
of pure epicyclic motion (equation  \ref{eqn:vfind}).
If the free epicycle is small ($a=0$) or the disk is cold  then 
\begin{equation}
{u \over v}
\left({ \Omega_g  \over m(\Omega - \Omega_p)} \right) \approx  - \tan m\phi_1
\end{equation} 
with
\begin{equation}
\phi_1 =  (\theta - \Omega_p t + \alpha \ln r_g + c_0)
\label{eqn:phi1}
\end{equation}
and is sensitive to the angle with respect to
the potential perturbation (compare to equation \ref{eqn:theta1}) for
pure epicyclic perturbations).  Likewise the distance from origin on the $u,v$ plane is 
dependent on the amplitude of radial oscillations
 with
\begin{equation}
 C(r_g) \approx \sqrt{{u^2 \Omega_g^2 \over m^2(\Omega_g - \Omega_p)^2 } + v^2  }
\end{equation}
 (compare to equation \ref{eqn:aepi}).

In the limit $\bar a < |C|$, particles are in
nearly elliptical orbits but the orientation of the ellipses can vary with radius.
A set of ellipses with differing orientations is shown
in Figure \ref{fig:cartoon}a. Each ellipse has the same eccentricity
but is shifted in angle with respect to the previous one.  The difference
in angle is 20$^\circ$ except between the 6th and 7th ellipse.
The difference in orientation angles between these two ellipses is
larger, 90$^\circ$, and introduces a discontinuity that we will discuss
later when we consider the relationship between gaps in the 
velocity distribution and discontinuities in the spiral structure.
The red circle on Figure \ref{fig:cartoon}a shows an example of an arm peak that
lies outside the discontinuity.
In this neighborhood there are nearly overlapping orbits.
The range of velocities seen
in this neighborhood would be high as there are stars 
from many different orbits passing through the neighborhood.
There should be a relation between $u$ and $v$ velocity components 
as there is a relation between mean galactic radius of the ellipse and 
orbital angle in the neighborhood.  Thus 
the velocity distribution would exhibit an arc.

The green circle in Figure \ref{fig:cartoon}a 
shows an interarm region where the angles of
orbits only varies slightly compared to that in 
the red neighborhood.  In the green interarm neighborhood we expect
a narrow velocity distribution.   This 
illustrates why we see expect to see arcs in the velocity distributions
on spiral arm peaks and lower velocity dispersions 
in interarm regions.

An arc in a local velocity distribution that has $u$ decreasing as $v$ increases can be interpreted
in terms of orbital properties as a function of $r_g$ 
(approximately set by $v$).
The positive $v$ orbits have larger guiding radii.
The change in angle or $v/u$ slope on 
the velocity distribution plots implies that the epicyclic angle $\phi_1$
(equation \ref{eqn:phi1}) decreases with
increasing $r_g$.  Thus these arcs correspond to a relation
between epicyclic angle and guiding radius.
Consider our cartoon, Figure \ref{fig:cartoon}a, showing concentric
ellipses with different orientation angles. 
A neighborhood on top of a spiral arm peak (such as the red one) intersects
orbits with a range of epicyclic angles,
corresponding on this figure to a smooth change in orientation of the ellipse.
Thus we expect that an arc would be observed in a velocity distribution
at the position of a spiral arm peak.
The angle of the arc in the velocity distribution
is likely to be affected by the winding angle of the 
spiral structure with a more open
spiral corresponding to a arc with steeper slope.  
However a neighborhood may contain orbits with a smaller range 
of guiding radii if the spiral structure is less tightly wound. 
Model orbital distributions would be needed to use
the slopes and ranges of velocities 
of clumps seen in velocity distributions, and their gradients,
to place constraints on the spiral structure (pattern speed,
pitch angle and strength) from observations of the velocity distribution
alone.

We now consider the type of arc that would be seen on a trailing
spiral arm. The red neighborhood in Figure \ref{fig:cartoon}a
intersects orbits near apocenter with small guiding radii (and so 
negative $v$) and
orbits that are near pericenter that have larger guiding radii (and so
positive $v$).
In between and in the neighborhood are orbits 
that are moving toward the Galactic
center and so have positive $u$ values.
For trailing spiral structure we expect the arc to progress from 
low $v$ to high $v$ passing through the top right quadrant of
the $uv$ plane, as we seen in arm peaks in Figures \ref{fig:page120},
\ref{fig:pages6} and \ref{fig:pages10}.  Thus our illustration Figure \ref{fig:cartoon}a
provides an explanation consistent with spiral
structure for the arcs seen in the velocity distribution.

We can use our first order perturbation model (equations \ref{eqn:rrg} - \ref{eqn:vrg})
to consider particles with nearby guiding radii.
We expand equation (\ref{eqn:rrg}) for closed orbits ($a=0$)
\begin{eqnarray}
r(r_g + x, \theta,t) &\approx & r_g + x   \\
 &&   +  C(r_g) \cos(m(\theta -\Omega_p t + \alpha \ln r_g + c_0 ))  \nonumber  \\
  && - C(r_g) \sin(m(\theta-\Omega_p t + \alpha \ln r_g + c_0)) {m\alpha x \over r_g} \nonumber
\end{eqnarray}
where we have used the WKB approximation by neglecting $dC(r_g)/dr$.
The derivative of the above equation
\begin{equation}
{dr \over dr_g} \approx 1 - C(r_g) \sin(m(\theta-\Omega_p t + \alpha \ln r_g +c_0)) {m\alpha  \over r_g}.
\end{equation}

Consider two closed orbits separated by a small difference in guiding radius.
The radial distance between these closed orbits at a particle angular location is smallest
when the above derivative is a minimum.
A small difference in current radius is possible when the sine term is negative.   This happens
midway between pericenter and apocenter.  Thus the closed orbits are nearest to each other
 when this sine term is negative ($m \phi_1 = \pi/2$).  The radial
velocity is also proportional to the sine of the angle.  For $\alpha >0$, corresponding
to trailing arms, we find that the orbits are closest together
when the radial velocity component is negative and the particles are approaching pericenter 
(positive $u$, zero $v$),
as shown by the red circle in Figure \ref{fig:cartoon}a.   Orbits are maximally
distant when the radial velocity is positive or 90 degrees away from the red circle
on Figure \ref{fig:cartoon}a that shows a two-armed structure.   At locations and corresponding velocities where the above derivative
 is high, the phase space density would be higher and so there would be features seen in a local velocity distribution.  Inclusion
of the the derivative $dC(r_g)/dr$ when estimating the derivative $dr/dr_g$ would shift the
estimated angle of the minimum of the derivative.    This shift may explain why arcs in 
the velocity distribution are primarily located in the upper right hand quadrant (positive $u,v$) and
not centered at $u>0$, $v=0$.

The above discussion has neglected the dispersion of orbits
about the closed or periodic ones shown in our cartoons
(Figure \ref{fig:cartoon}), however if the dispersion is neglected
then the predicted velocity distribution contains particles at
only a few velocities.
The stellar dispersion has been used in constructing weights 
for orbits to populate the velocity histograms and predict
velocity distributions from test particle integrations         
(e.g., \citealt{dehnen00,quillen05}). 
We should remember that Figure \ref{fig:cartoon} only
qualitatively gives us a relation between orbital and
velocity distribution, however
the properties of closed orbits are relevant for interpretation
of these distributions (e.g, \citealt{quillen05}).

The above equations are given in the WKB approximation for a spiral perturbation.  
For a bar or two-armed open spiral perturbation to first order the orbits are also ellipses however the 
WKB approximation cannot be used.  For a bar perturbation an additional term in $v_\theta$ should be taken into 
account when estimating the guiding radius  and the angle $m(\theta-\Omega_p t)$ 
from $u,v$ (see \citealt{B+T} section 3.3).   For a spiral perturbation if the WKB approximation is not used
additional terms in both velocity components would be used to estimate the guiding radius, the
and orbit orientation and the angle with respect to it from $u,v$  (e.g., \citealt{cont75}).

\section{Gaps at different radii}

We now look at the location of gaps in the velocity distribution
for the snapshot shown in Figures \ref{fig:page120} and \ref{fig:dcirc120}
at $T = 600$ Myr.
In Figure \ref{fig:page120} at a radius of $r_0 = 5.1$ kpc and an angle
of $\phi_0 = 135^\circ$ there are two gaps seen in the velocity distribution,
one centered at $v\sim 50$ and the other at $v\sim 0$ km/s.  The one
at $v\sim 0$ has a guiding radius of $r_g \sim 5.1$ kpc, 
the same as the neighborhood radius.  
Using equation \ref{eqn:vfind}, we can predict the $v$ value
for this gap were it to be observed from 
at a neighborhood radius of $r_0 = 4.1$ kpc.  We estimate that it should
lie at $v \sim $ 50 km/s.  There is
an arc with a gap at about $v \sim 50$ km/s in the velocity distribution 
at $r_0 = 4.1$ kpc and at $\phi_0 \ = 135^\circ$, as expected.
The other gap in the neighborhood with $r_0 = 5.1, \phi_0 = 135^\circ$ 
at $v \sim 50$ km/s has a guiding radius $r_g \sim 6.3$ kpc (again using
equation \ref{eqn:vfind}).  A clump with this
guiding radius would lie at $v \sim 0$ km/s in a neighborhood
with $r_0 = 6.4$ kpc.
We don't see a gap with this $v$ value at $r_0 = 6.4$~kpc and
$\phi_0 = 135^\circ$ but we do see one in 
the neighboring position at $r_0 = 6.4$ kpc, $\phi_0 = 90^\circ$.
Thus we find that relation between $v$ and guiding radius is consistent
with shifts in individual features in $v$ as a function of
neighborhood radius.

We also find gaps at nearly zero $v$ values in neighborhoods with
$r_0 \sim 8$ and 10 kpc.
In the inner galaxy
there is a gap at $r_0 = 4.1$ kpc, $\phi_0 = 45,90^\circ$
with slightly negative $v$
that is also present at $v \sim -50$ km/s in the neighborhood
with $r_0 = 5.1$ kpc, $\phi_0 = 45^\circ$ and has $r_g \sim 3.9$ kpc.
Altogether we see gaps at the following guiding radii
$r_g \sim$ 3.9, 5.1, 6.3, 8 and 10 kpc at this time in the simulation.

Looking back at Figure \ref{fig:page120} we compare the location of gaps
in the velocity distribution (identified by their guiding radii)
and the locations of kinks (changes in winding angle) or discontinuities
in the spiral arms.  In Figure \ref{fig:dcirc120}b we have identified
radii at which we find discontinuities in the spiral structure,
either changes in winding angle (kinks) or places where spiral arms
or armlets appear or disappear.
These radii are shown as black 
circles in Figure \ref{fig:dcirc120}a and as
horizontal lines in figure \ref{fig:dcirc120}b.
The radii of the discontinuities on these plots are 
at $r_d \sim 3.5$, 5.1, 6.3, 8.7 and 11 kpc. 
A comparison between the list of discontinuities and our list
of gaps identified in the velocity distributions (identified by
their guiding radii) suggests
that these two lists are related.

A gap in the velocity distribution occurs in a neighborhood when there are two groups
of stars each with velocity vectors in similar but different directions, 
but few stars with velocity vectors in between those of the two groups.  
In Figure \ref{fig:cartoon}a we show concentric ellipses that increase in radius
and shift in orientation angle.  The difference in orientation angle
between each ellipse is small and constant except 
between the 6-th and 7-th ellipse where the difference is much large 
and 90$^\circ$.  There is a discontinuity in the 
position of the spiral arm peak at the radius
of the introduced large change in orientation angle 
that we can see from the shape of the orbital
overlap regions.   The discontinuity introduces
regions such as the neighborhood illustrated with the blue circle.  This
neighborhood contains orbits with two separate groups of orbit angles. 
In this neighborhood we would expect a corresponding gap in the velocity distribution.
Figure \ref{fig:cartoon}a shows that we expect gaps in the velocity
distribution where there are discontinuities in the spiral structure, and 
suggests that gaps in the velocity distribution only occur at radii
where there are discontinuities in spiral structure and at specific angles where
there are spiral arm peaks.  We don't see gaps in every
velocity distribution in Figure \ref{fig:page120} but do
tend to see them where the surface density is high.
This is expected as density peaks can occur when multiple pattern peaks
coincide.

In the previous section  we introduced a first order perturbative orbital model containing a single
potential perturbation (equations \ref{eqn:rrg} - \ref{eqn:vrg}).  
This model can be extended for multiple patterns giving orbits with
radius, radial and tangential velocity as a function of guiding or mean radius to first
order in the perturbation strengths
\begin{eqnarray}
r(r_g,\theta, t) &=& r_g + a \cos (\kappa_g t + \phi_0)   \label{eqn:mrg} \\
                       && +C_1(r_g) \cos (m_1(\theta - \Omega_1 t + \alpha_1 \ln r_g + c_1) \nonumber \\ 
                       && + C_2(r_g) \cos (m_2(\theta - \Omega_2 t + \alpha_2 \ln r_g + c_2) \nonumber 
\end{eqnarray}
When $\bar a $ is small compared to $|C_1|$ and $|C_2|$ the orbits
are nearly closed as shown in Figure \ref{fig:cartoon}a,b.

For a particular neighborhood with  $r,\theta$ the above equations must be inverted
to solve for $r_g$, $a$ and $\phi_0$ as a function of the velocity components.
The inversion need not give a unique value of $r_g$ so particles from separated radial
regions (in guiding radius $r_g$) can simultaneously be present in one local neighborhood.
The similarity of terms in $r, u, v_\theta$ imply that there is a close
relation between the inverted $r_g$ distribution and the velocity components.
This situation is illustrated in Figures \ref{fig:cartoon}a,b where
 transitions between one pattern and another gives neighborhoods containing
orbits from different regions of the galaxy and a bifurcated velocity distribution in that
neighborhood.  Figure \ref{fig:cartoon}a shows a transition between one two armed
and another two-armed pattern.   Figure \ref{fig:cartoon}b illustrates a transition
between a two-armed and a three-armed pattern.  In this figure 
the inner 6 curves are ellipses however the outer ones
are a sum of elliptical and triangle perturbations such
as might occur if there are simultaneously two-armed and 
three-armed spiral density waves.  The elliptical and triangular
perturbations vary smoothly in orientation angle from curve to curve. 
The transition region exhibits discontinuities in overlap regions caused
by the onset of the three-armed pattern.
The blue circle is located at a radius where there is a discontinuity
and shows that the angular distribution of orbits is bifurcated.  We expect that
in such a region there will be gap in the velocity distribution.

A neighborhood with $r,\theta$ is most likely to contain two separate solutions for $r_g$
if the phases of the arguments for each perturbation have opposite signs.
For example if the angle $m_1(\theta - \Omega_1 t + \alpha_1 \ln r_g + c_1) \sim 0$ and $C_1>0$ then
$r>r_g$ and the particles in a neighborhood at $r$ will have guiding radii interior to $r$.
The orbits will have $u \sim 0$ but $v$ greater than the circular velocity at $r$ (consistent
with the discussion in section 5.5).
If the angle $m_2(\theta - \Omega_2 t + \alpha_2 \ln r_g +c_2) \sim \pi$ then the opposite is true.  A neighborhood
at $r$ can simultaneously contain stars on orbits associated with an interior perturbation
near apocenter and stars associated with an outer perturbation near peri-center. 

Our spectrograms showed that there are both two-armed and three-armed
patterns present in this simulation.  The two-armed and 
three-armed patterns have different radial extents and different pattern
speeds.   Figure \ref{fig:cartoon}b shows that the transition
between a two-armed and three-armed pattern can  cause a discontinuity
in the spiral arm peaks.  This type of discontinuity can 
also give neighborhoods with two sets of orbital orientation angles
and so gaps in a velocity distribution, as illustrated by the
blue neighborhood in this figure.

We now compare the list of radial discontinuities and gaps with estimated
resonance locations
identified from the pattern speeds in the simulations.
If we only had one strong wave then we would be restricted
to a few resonances.    However our simulation shows
multiple patterns,  each with a different radial range.   
Keeping this in mind, perhaps it is not surprising that the
density distribution shows multiple spiral arm discontinuities
and the velocity distributions show multiple gaps.

If the spiral patterns are excited at resonances then
we expect that they will start or/and end near resonances.
In section 4 we discussed likely ranges for these patterns
with likely resonances as boundaries.
The innermost discontinuity, seen both as gaps in the
velocity distribution and in the spiral morphology, lies at radius
of about $r_g \sim 4$ kpc.  This lies near the bar's $m=4$ ILR and 
is not distant from the $m=2$ ILR of the inner two-armed
spiral.  The discontinuity at $r_g \sim 5$ kpc lies near
the bar's corotation resonance, the 
the $m=3$ ILR of the three-armed spiral and is not distant
from the $m=4$ ILR of the inner two-armed spiral and the $m=2$ ILR
of the outer two-armed spiral.   
The discontinuity at a radius $r_d \sim 11$ is near the 
$m=3$ OLR of the three-armed spiral and the $m=2$ OLR of the inner
two-armed spiral.  
That at $r_\sim 8$ kpc is near the bar's $m=2$ OLR and the $m=2$ ILR
of the outer two-armed spiral.  
It also may be the outer boundary
of the inner two-armed spiral and near its $m=4$ OLR.
The division at $r_d \sim 6.3$ is not obviously associated with
a resonance but could be due to a division between the inner and outer
two-armed spirals.

The first order approximation used here has a small divisor
problem nearing resonance.   To understand the dynamics near resonance a higher
order model is required \citep{cont75}.  Approaching resonance
the epicyclic amplitude increases but does not go to infinity and closed orbits
exist on both sides of resonance.  Consequently a closed orbit approximation
(such as illustrated in Figure \ref{fig:cartoon})
may not be invalid even quite close to resonance.  
As epicyclic amplitudes become high near resonance, Lindblad resonances 
are particularly likely locations where orbits can cross into nearby neighborhoods.
Because the orientation of closed orbits can vary on either side of a Lindblad resonance they 
can also induce a gap in a local velocity distribution \citep{dehnen00}, also see Figure 4 by \citet{quillen05}.
If there are two perturbations, localized regions near resonance can become chaotic \citep{quillen03}
perhaps leading to the truncation of spiral arms.
The Hamiltonian model by \citet{quillen03} showed that periodic orbits can exist near resonance
even when two patterns are present though if there is strong resonance overlap
there may be regions lacking nearly circular orbits where
localized diffusion can be rapid \citep{shev11} and heating is possible \citep{minchev06}. 

The gaps in the velocity distribution and discontinuities 
in spiral morphology seen in our simulation can be explained with a model containing
multiple spiral density waves. The onset radii of
each wave introduces the discontinuities in the spiral arm morphology 
and associated gaps in the velocity distribution.
Our simulation shows multiple patterns that are coupled, and there 
are radii that are likely resonant with more than one pattern.
These radii are likely associated with onset radii
of waves, gaps in the velocity
distribution and discontinuities in the spiral arm morphology.

\subsection{Clumps like the Hercules Stream}

When the neighborhood lies on a discontinuity, two sets
of orbits can pass through the neighborhood;
see for example the blue neighborhood in Figure \ref{fig:cartoon}a.
In this case the first set is those associated
the spiral arm peak and their velocities would lie on an arc passing through the
top right quadrant in the $uv$ plane.
The other set of orbits comes from
the inner galaxy and differs in orientation by about 90$^\circ$.
This difference in orientation implies that its $u$ values will
have the opposite sign, and since the orbits have lower guiding
radii, they should have lower $v$ values.
Thus we expect the velocities of stars on the second group
of orbits can have low $v$ and the opposite sign of $u$ and so
form a clump on the lower left 
quadrant in the $uv$ plane, similar in position
to the Hercules stream at $v\sim -50$~km/s in our galaxy \citep{dehnen98}.

The $m=2$ outer Lindblad resonance with the bar at earlier times in the
simulation is at about 9 kpc.  If the solar neighborhood 
lies just outside this resonance with the bar in 
the Galaxy then the radius in this
simulation most matching our position in the Galaxy (just outside
the bar's OLR) is at $r_0 \sim 10$ kpc.  We show velocity distributions in 
neighborhoods with this radius and at different  times
in Figure \ref{fig:pages10}.
At this radius, clumps similar in position to the Hercules
stream are seen at angles (with respect to the bar)
of $\phi_0 = $ 90, 135, 270 and 315$^\circ$ but not at all times.
It is rare to see these streams at angles nearly aligned or perpendicular to the bar.
This suggests that non-circular motions caused by the bar 
may be needed to produce the large epicyclic amplitudes
corresponding to the strong negative $v$ values in the clump.  

The orientation of a periodic orbit can depend upon its location.
For example, orbits can be aligned with a bar
outside the $m=2$ Outer Lindblad resonance and aligned perpendicular
to it between this resonance and the corotation resonance.
A flips in orbit orientation on either side of a resonance
can also cause a gap in the velocity distribution
(e.g., \citealt{quillen05}).  
However if the orbital distribution was primarily determined
by the bar then we might expect the outer galaxy to look like
a ring galaxy.  However our simulated galaxy at $r_0 \sim 10$ kpc
has a strong two-armed spiral structure.  The bar's $m=2$ outer Lindblad 
resonance itself may also be a site for coupling between waves.
Both the bar and local spiral structure 
affect the Hercules stream like clumps 
in the orbital distribution.

The velocity distribution in the solar neighborhood not only displays a stream
at $v\sim -50$ km/s (the Hercules stream) but 
is triangular shaped for velocities with low $u,v$
We show in Figure \ref{fig:r10_sn} rightmost panel  the velocity distribution of the magnitude-complete, kinematically unbiased sample of 16682 nearby F and G dwarfs by 
\citet{holmberg09}.
A triangle shaped distribution at lower $u,v$ and a Hercules
stream like feature at negative $v$ are sometimes seen in our
velocity distributions at $r_0 = 10$ kpc, 
near the bar's $m=2$ outer Lindblad resonance, are both
seen for example the rightmost panel of Figure \ref{fig:comp}.
We have looked through our velocity distributions at this $r_0$ 
and with a bar orientation consistent with the $\sim 45^\circ$ angle (between bar semi-major
axis and Sun the Galactic center line) 
estimated from the SPITZER/GLIMPSE survey \citep{benjamin05} 
for other examples of velocity distributions that exhibit the gross features
present in the solar neighborhood velocity distribution.
These are shown in Figure \ref{fig:r10_sn} for smaller neighborhoods
with size set by $f=0.05$ along with disk surface density at the same timesteps
in Figure \ref{fig:sol}.  
In Figure \ref{fig:sol} blue circles show the location of neighborhoods
used to extract the velocity distributions.  The disk has been flipped so
that rotation is clockwise rather than counter clockwise and
rotated so the extraction neighborhood is oriented at the top
of the plot.  This has been done so that these images
can more readily be compared with models for the Galaxy.
The location of the mock Sun is at the blue circle with a galactic
longitude of 90$^\circ$ viewing to the right.

At certain times and locations in our simulation 
the local velocity distribution is similar to that seen 
in the solar neighborhood at a position
consistent with the orientation, length and distance of the Galactic bar.
By similar we mean with a triangular shaped feature at low $uv$ and a Hercules-like
stream at low $v$.
Weak clumps at low $u,v$ are seen in our velocity distributions, similar to
those labelled as low velocity streams such as the Hyades and Pleiades
stream in the solar neighborhood \citep{dehnen98}.  We do not match each of the low
velocity streams seen in the solar neighborhood velocity distribution but
our velocity distributions do exhibit low velocity clumps suggesting that they could be
related to spiral waves in the disk.    However none the models
have have discussed above would easily account for this structure.
Our Hercules like stream 
is more heavily populated by stars than that in the solar neighborhood, probably because our disk is hotter.
Our simulated Hercules stream
is at lower $v$ than that in the solar neighborhood suggesting that a somewhat smaller galactocentric radius would provide a better
analog, however at smaller radii it is more difficult to see a strong Hercules like stream in
the simulated velocity distributions.    

It is interesting to compare the spiral morphologies  seen in Figure \ref{fig:sol}.
In all cases the location of the mock Sun is just exterior to a strong spiral arm.  
This spiral arm would have tangent (as viewed from the mock Sun)
at a Galactic longitude of about $-50^\circ$ (to the left on the plot), 
and approximately consistent with 
the location of the Centaurus Arm tangent measured by \citet{churchwell09}
from the GLIMPSE survey counts
and denoted as the Crux-Scutum Arm tangent by \citet{vallee08}.
Our simulated galaxy would exhibit only one or two peaks or tangent arms in a projected
number count diagram (similar to figure 14 by \citealt{churchwell09})
and so would resemble the observed infrared observations.
As our simulation lacks gas we cannot see additional gas rich but star poor
spiral arms that are likely present in our Galaxy \citep{drimmel01}.
While all the images in Figure \ref{fig:sol} have a strong spiral arm just
interior to the mock Sun, the opposite side of the galaxy has a variety of
morphologies.  This implies that the solar neighborhood velocity distribution
alone is not sufficient to break the degeneracies of a model 
that includes multiple spiral waves.

\section{Summary and Discussion}

Using an N-body hybrid simulation we have looked for a relationship 
between structure in local velocity distributions and spiral and bar
density waves.  Patterns for these waves are identified
through Fourier spectrograms. Previous studies (e.g., \citealt{rau99}),
have suggested that there are couplings between waves.
The similarity between spectrograms at different stages in 
our simulations (as the bar slows down) suggests that
our simulation also exhibits couplings between waves,
with bar and lopsided waves likely driving spiral waves.
Even though multiple waves are present, they could be coupled and slowly
evolving rather than transient.

We find that arcs in the local velocity distribution are seen in neighborhoods
located on arm peaks, whereas interarm regions tend to have lower velocity 
dispersions.  This behavior is understood by 
considering a series of nested elliptical orbits, appropriate
for a cold disk influenced by a single spiral perturbation.   Spiral arm peaks
are present where ellipses touch or overlap and 
these locations correspond to neighborhoods
containing orbits with a wider range of velocity vector orientations than
interarm regions.

Gaps can be seen in the velocity distributions in our simulation all over
the simulated galaxy, confirming the early finding by \citet{fux00}.  However,
the $v$ values of these gaps correspond to a small number of guiding radii 
that are near resonances with the bar and spiral arm patterns  
or where there are kinks or discontinuities in the spiral arm morphology.
The relation between gaps in the velocity distribution and
discontinuities in spiral structure is understood by
considering a series of nested ellipses that have a discontinuity
in their relative orientation angles.  The discontinuity introduces
neighborhoods containing overlapping orbits that have a bifurcated
distribution of orbit orientations. This bifurcated distribution
corresponds to two clumps and a gap separating them in the 
velocity distribution.
Multiple waves, each with a different pattern speed, present in the 
simulation, introduce radii at which there are discontinuities
in the spiral structure.
Consequently gaps in the velocity distributions can be seen at
multiple radii in the disk.

N-body simulations have the advantage over test particle simulations
that the orbits of particles are consistent with the gravitational potential
generated by the particles themselves. 
Additionally N-body simulations naturally
contain varying or evolving bar and spiral
structures that would require additional and possibly poorly constrained
parameters to describe inside  test particle simulations.
The study here, though preliminary and based on only one hybrid simulation,
can be considered to be complimentary to studies
carried out with test particle simulations.

\subsection{Implications for Studies of the Milky Way}

From our simulations
we have found that interarm regions tend to have lower velocity dispersions
than arm peaks. They also 
tend to lack arcs and gaps in the velocity distribution.
The Sun is likely in an interarm region \citep{vallee08}.
We may have biased view of the Milky Way disk dynamics
because the local velocity distribution does not represent that
at other positions in the Galaxy.
For example the mean local velocity dispersion in the Galactic
disk may have been underestimated because of our 
interarm location in the Galaxy.
Large arcs in the local stellar  velocity distribution may be
present in nearby spiral arm peaks.

There remains little consensus among different studies for 
the pattern speed of local spiral structure 
(e.g. \citealt{shaviv03,lepine10,gerhard10}).  The lack 
of consensus may arise because 
the presence of more one wave complicates 
analysis (e.g., \citealt{naoz07}).
The Local Spur which is thought to lie between the dominant
two-armed or four-armed structures may arise because
multiple waves are passing in and out of
phase with each other (e.g., \citealt{henry03,meidt09}).
As armlets can appear and disappear, a localized short burst of star
formation may occur that does not continue to progress around 
the galaxy as is expected when a sole wave passes through
the disk. Evidence for short isolated
bursts of star formation in the disk would support interpretation
of the Galactic disk in terms of multiple waves simultaneously present in the disk.
We have noted that density peaks due to
constructive interference of patterns tend to begin at smaller radius
and progress to larger radius before fading away.  
Trails of recently formed clusters with age decreasing with increasing radius
may be found in the Galaxy disk.

Recent studies have placed strong constraints on the 
pattern speed of the Galactic bar from the morphology
of the Hercules stream  in the local velocity distribution
\citep{dehnen00,minchev07,gardner10}.
The association of the $v$ value of the gap separating
the Hercules stream from the rest of the stars with the bar's 
$m=2$ outer Lindblad Resonance tightly constrains the bar pattern speed.
However features similar to the Hercules stream are seen in our
velocity distributions that both associated with the bar's $m=2$
outer Lindblad resonance and at other locations in the galaxy. 
These clumps do not remain fixed in time and are affected
by the coupled spiral structure in the outer Galaxy.
Previous models of the Hercules stream
have neglected spiral structure and have been 
based on test particle simulations, not N-body simulations 
(with the exception of the study by \citealt{fux00}).
If wave coupling is common then 
the bar's outer Lindblad resonance may be coincident with
a Lindblad resonance  with a spiral pattern.
Future models could consider both spiral and bar patterns when interpreting
the velocity distribution near the bar's outer Lindblad resonance.

Velocity distributions with morphology similar to that in the solar neighborhood
are seen in our simulation at location just exterior to the bar's $m=2$
outer Lindblad resonance with bar orientation consistent with infrared
studies.    By similar we mean, showing a Hercules like stream and a triangular
shape at lower $u,v$.
The velocity distribution more closely resembles that seen in the solar neighborhood
when there is a strong spiral arm,  consistent with the observed Centaurus Arm tangent,
just interior to the solar neighborhood.

Like many other galaxies, the Milky Way is lopsided \citep{levine06}. If coupling
between  a lopsided motion and a Galactic bar is common then we might expect a strong
three armed spiral wave would be present in the solar neighborhood.   However, almost all
studies have assumed two or four armed spiral structures \citep{vallee08}.  If a strong
three armed wave is present in the solar neighborhood and its inner Lindblad resonance
is coincident with the Galactic bar's corotation radius (as likely true in our simulation) 
then we can estimate its pattern speed.   
At a inner Lindblad resonance the pattern speed is related to
the angular rotation rate by  $\Omega_p = \Omega(1 - {\kappa \over m\Omega})$.
If we set $\Omega$ in this relation equal to the bar pattern speed (approximate 1.9 times the angular
rotation rate at the solar circle $\Omega_\odot$; \citealt{gardner10}), use
a flat rotation curve with ${\kappa \over \Omega}  = \sqrt{2}$ and $m=3$ we estimate
a pattern speed of $\Omega_3 = (1 - \sqrt{2}/3) 1.9 \Omega_\odot \approx \Omega_\odot $
putting the Sun near corotation of this three armed pattern.  
Pattern speeds for spiral structure fall into two groups, a faster set, placing the Sun
near the corotation resonance and a slower set placing the Sun near the inner 4:1 Lindblad resonance
(see Table 3 by \citealt{shaviv03}).
A three armed pattern with pattern speed similar to $\Omega_\odot$ would
have $m=3$ inner Lindblad resonance near the bar's corotation resonance and
so could be coupled to the bar. 
In our simulation a two spiral pattern is also prominent near the position of the mock solar
neighborhood with inner Lindblad resonance
near the bar's corotation rotation resonance.  We can estimating the pattern
speed of a two armed wave satisfying this condition from the bar's pattern speed using
$m=2$ finding a pattern speed of $\Omega_2 \approx 0.55 \Omega_0$ or about
14 km s$^{-1}$ kpc$^{-1}$ consistent with a slower group of pattern speeds estimated
a number of studies (see Table 3 by \citealt{shaviv03}).

\citet{lepine10} propose that a nearby kink in a spiral arm winding angle 
is associated with a single Lindblad resonance with a spiral pattern.  
However our simulations
suggest that kinks and discontinuities in the spiral structure
not only are associated with Lindblad resonances with a single pattern
(as also explored by \citealt{quillen05,cont88}) but are also
signatures of transitions from one pattern to another.  Rather than
rare, our simulation suggests that these discontinuities may exist
all over the Galaxy. 

The mode or wave coupling possibly present in our simulations implies that resonances
between different patterns could be commonly coincident. As shown
by \citet{minchev06} this causes chaotic
motions in the stellar orbital motions and so accelerated heating
(or increasing velocity dispersion) near localized regions of resonance overlap.
If waves are coupled in the Milky Way disk then we might
except both localized accelerated heating and 
radial migration \citep{minchev10a,minchev11} as a consequence
(also see \citealt{quillen03,voglis06,shev11}).

The waves present in our simulation are likely coupled and so not transient. 
The divisions seen in the velocity distribution we interpret in terms
of discontinuities associated with resonances and transitions
from one pattern to another rather than a model
involving stochastic growth of transient spiral structure 
(e.g., \citealt{desimone04}).

\subsection{Future Work}

This study only considered one single simulation and it ran
only for 1.3 Gyr.
The simulation is not ideal as we have undersampled the bulge
and halo in order to well resolve the disk, and the energy
conservation is not excellent.
However, this particular simulation is particularly rich in spiral density
waves and so a good one for comparing velocity distributions
to structure seen in spectrograms.
While our galaxy was simulated in isolation, real galaxies suffer 
external perturbations.   Lopsidedness in our simulation that may
be an artifact of our undersampled live halo, is commonly seen in external
galaxies \citep{jog09}. Hence the coupling between bar, lopsided motion
and three armed waves that we suspect is present in our simulation 
may be ubiquitous rather than just a curiosity.

In our simulations we purposely undersampled the halo and bulge
particle distributions so that we could better resolve the galaxy disk.  
However as a result our few and relatively massive halo particles 
cause spurious heating in the disk.  
We have measured the radial velocity dispersion in the disk at a radius of 10 kpc
and found that it does not increase by more than 10 km/s across the entire simulation.   
This suggests that heating by massive halo particles 
is not excessive in this simulation.  The similarity between spectrograms in the middle
and later part of the simulation suggests that waves are coupled to the bar
rather than randomly generated by noise associated with the granularity
of our halo during the simulation.
Were we to raise the number of massive particles in the disk, halo and bulge, 
the simulations would exhibit reduced numerical 
noise and so might have weaker or delayed spiral and bar structures.
Future studies should not only consider more accurate simulations with
larger numbers of particles but also explore simulations with
differently distributed disks, bulges and halos, 
the role of a cold gaseous component and perturbations due to mergers.

\begin{figure*} 
\includegraphics[angle=0,width=6.5in]{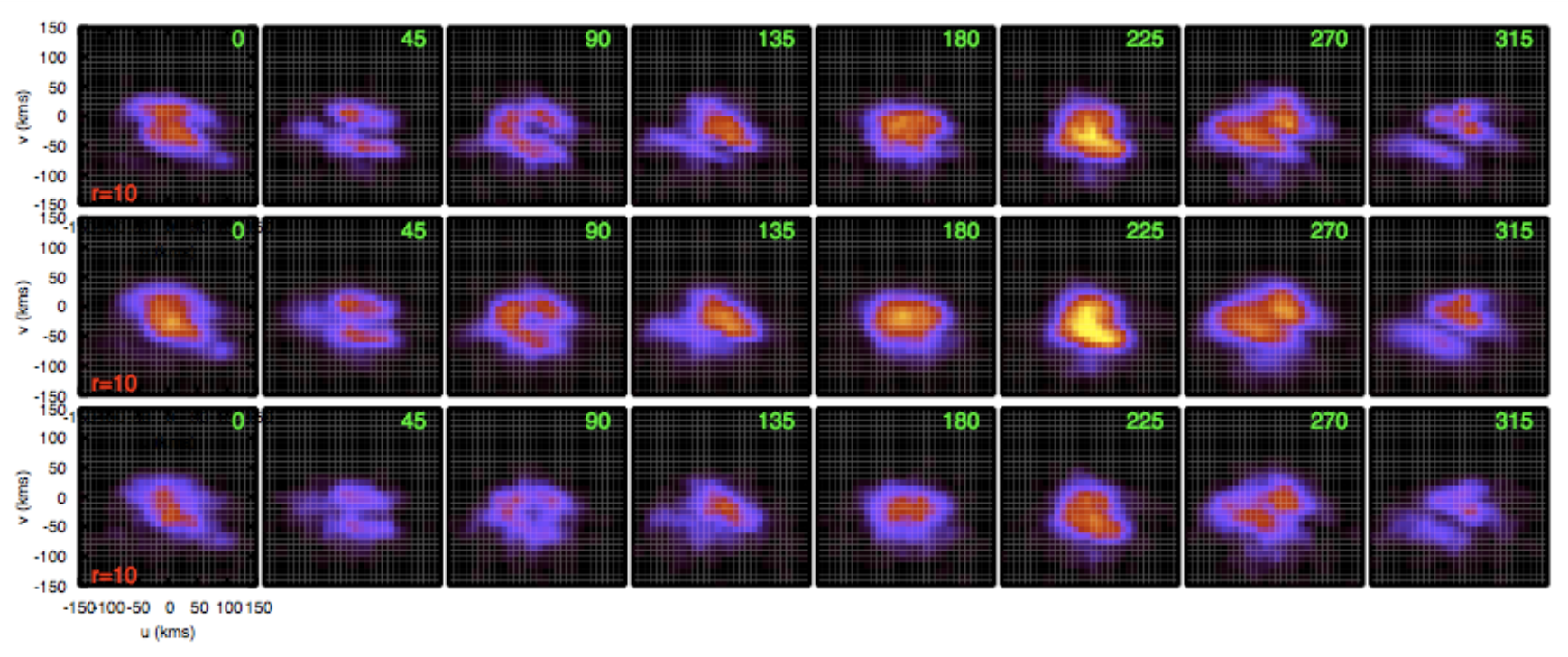}  
\caption{
A comparison between velocity distributions extracted
at a radius of $r_0 = 10$ kpc using neighborhoods with radius with
500 pc ($f=0.05$; top panels) and  1000 pc ($f=0.1$; middle row of panels).
The simulation time was arbitrarily chosen and is 800 Myr.
The bottom row is similar to the middle panel but shows only massive
disk particles whereas the distributions in the top two rows are constructed
from both massive and tracer particles.  
The neighborhoods have the same angles as in Figure \ref{fig:page120}
and are arranged in the same way.
When the neighborhoods are larger the velocity distributions are smoother
and less grainy as would be expected because more particles are found
in each neighborhood.
\label{fig:comp}
}
\end{figure*}

\begin{figure*} 
\begin{center}
$\begin{array}{cc}
\includegraphics[angle=0,width=3.5in]{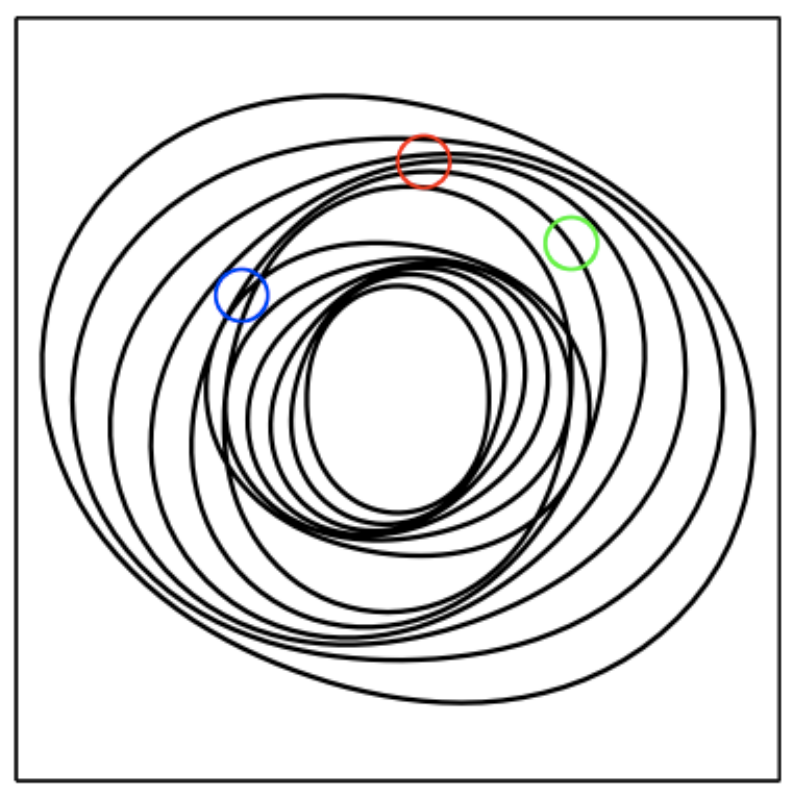}  &
\includegraphics[angle=0,width=3.5in]{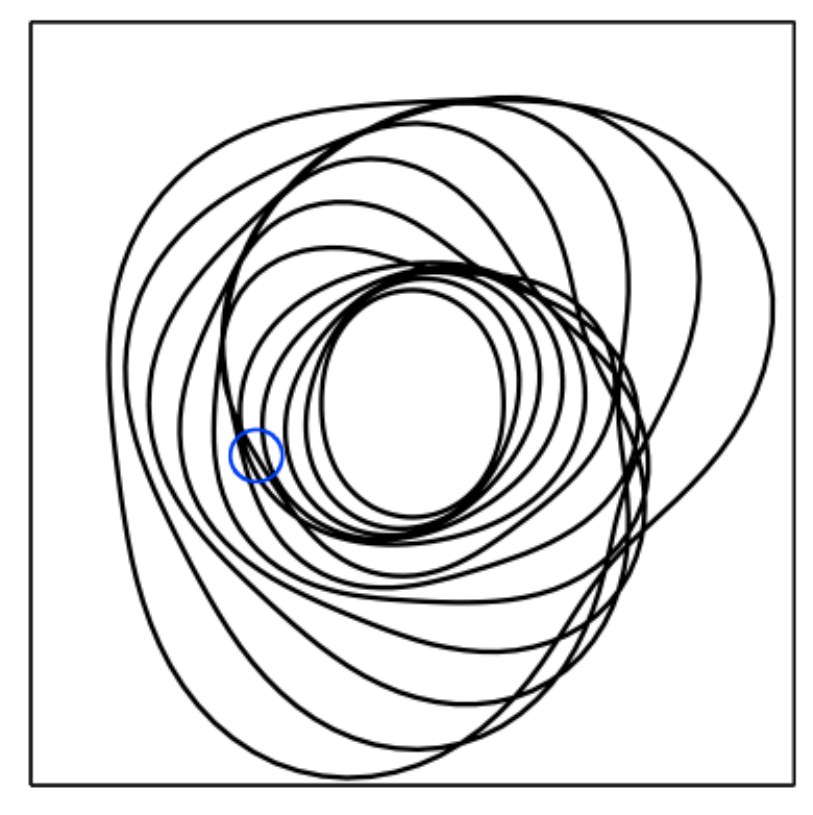} 
\end{array}$
\end{center}
\caption{
a) A series of ellipses is shown with semi-major axes
logarithmically spaced from the other.
For a cold disk, to first order in the perturbation strength and for
an $m=2$ perturbation, orbits are approximately ellipses.
The orientation of each ellipse semi-major axis is 20 degrees counter-clockwise
lower than that just inside it.  The only exception to 
this rule is between the 6-th and 7-th ellipse (counting
from smallest to largest) where the angular
difference is larger.    Spiral arm peaks (density peaks
in a stellar distribution) occur where the orbits are close together or overlap. 
The red circle is an example of an arm peak. 
There are a large range of velocities or orbit orientations in a region where
many elliptical orbits overlap, giving an arc in the velocity distribution.
In interarm regions, (e.g., green circle) 
the range of velocities (or angles on this plot)
in a particular neighborhood is low and so the corresponding 
velocity dispersion in this neighborhood is also low.
The spiral arms are trailing with galactic rotation
in the counter clockwise direction. 
On an arm a broad range of angles exists in the orbits crossing
the neighborhood, so the arc in the corresponding velocity space is continuous.
The angular offset between the 6-th and 7-th ellipse, as might occur
if there are different patterns in different regions of the disk,  causes
a discontinuity in the overlap region that manifests 
as a gap in the velocity distribution.
b) The inner 6 curves are ellipses.  The outer ones
are a sum of elliptical and triangle perturbations such
as might occur if there are simultaneously two-armed and 
three-armed spiral density waves in different regions of the disk.  
The transition region exhibits discontinuities in overlap regions caused
by the onset of the three-armed pattern.
The blue circle is located at a radius where there is a discontinuity
and shows that the angular distribution of orbits is bifurcated.  We expect that
in such a region there will be gap in the velocity distribution.
\label{fig:cartoon}
}
\end{figure*}

\begin{figure*}  
\includegraphics[angle=0,width=6.5in]{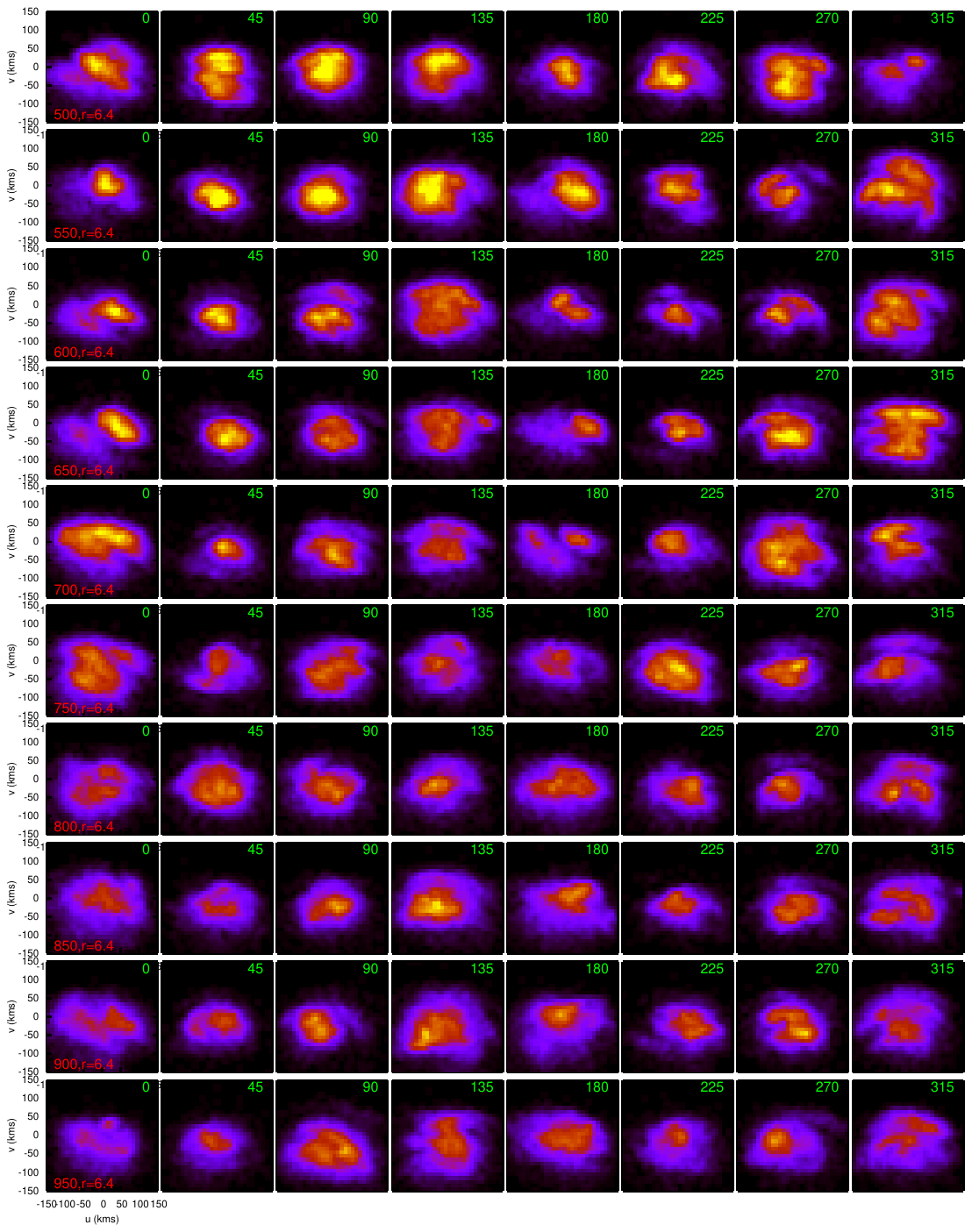} 
\caption{
We show $uv$ velocity distribution histograms generated with neighborhood
radii of $r_0 = 6.4$~kpc but at different times.  
From top to bottom row the histograms were generated from
snapshots at output times 500 to 950 Myrs.  Each snapshot is separated in time
by 50 Myr. As in Figure \ref{fig:page120} the angles
are given with respect to the bar and the same sequence of angles is shown.
Gaps in the velocity distribution tend to remain at the same $v$ values.
Even though the positions shown are fixed in the bar's rotating frame 
there are variations in the features seen in the velocity distribution
that must be due to the spiral waves in the outer disk.
\label{fig:pages6}
}
\end{figure*}

\begin{figure*} 
\includegraphics[angle=0,width=6.5in]{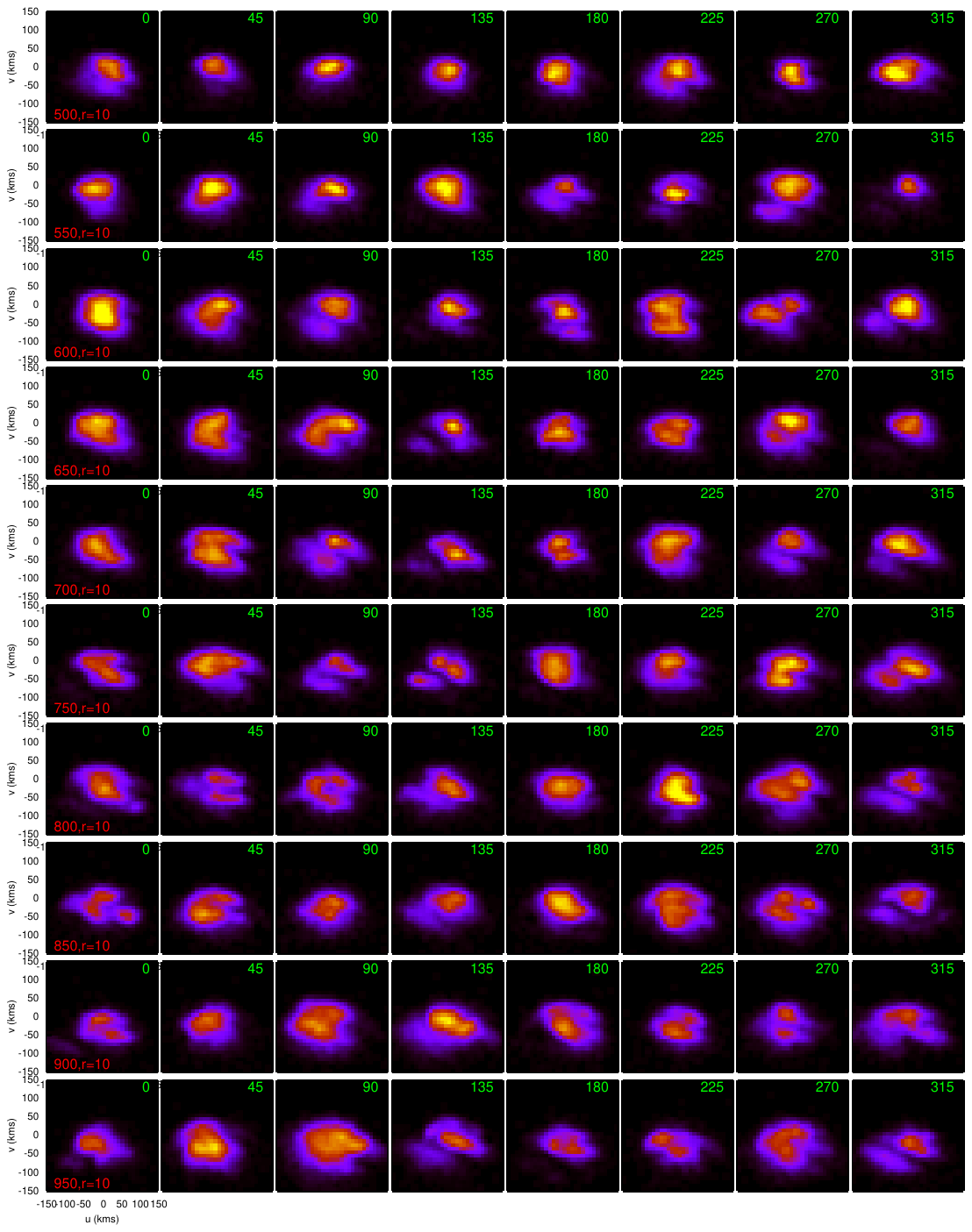} 
\caption{
Velocity distribution histograms generated with neighborhood
radii of $r_0 = 10.0$~kpc but at different times.  
Similar to figure \ref{fig:pages6}.
These neighborhoods lie just outside the bar's $m=2$ 
outer Lindblad resonance (at about 9 kpc).
A clump similar to the solar neighborhood's Hercules stream (with both negative $u$ and $v$) 
is primarily seen 
at angles $\phi_0 = $ 90, 135, 270 and 315$^\circ$ respect to the bar. 
It is rare to see these streams at other angles with respect to the bar.
Non-circular motions caused by the bar 
may be required to induce the large epicyclic amplitude
implied by the low $v$ values in the clump.
\label{fig:pages10}
}
\end{figure*}

\begin{figure*} 
\includegraphics[angle=0,width=7.0in]{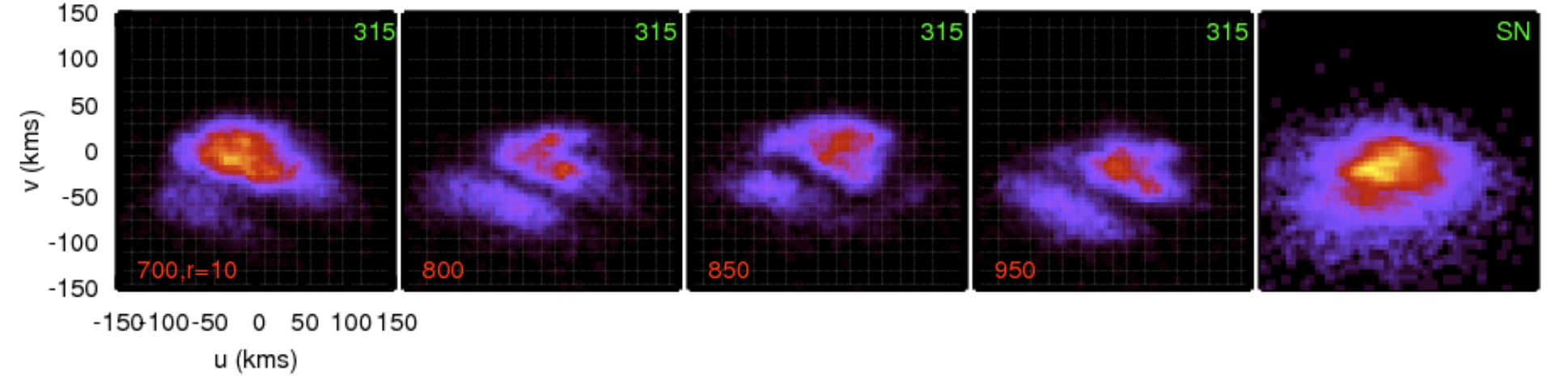} 
\caption{
The leftmost 4 panels show
velocity distributions at a radius of $r_0=10$ kpc chosen because they exhibit
both a Hercules-like stream and a triangular shape near the origin similar
to the  morphology of the solar neighborhood's velocity distribution (shown on the right).
These velocity distributions
where created with neighborhoods with $f=0.05$.  The time (in Myr) of each snapshot is on the lower left of each panel.
\label{fig:r10_sn}
}
\end{figure*}

\begin{figure*}  
\includegraphics[angle=0,width=7.0in]{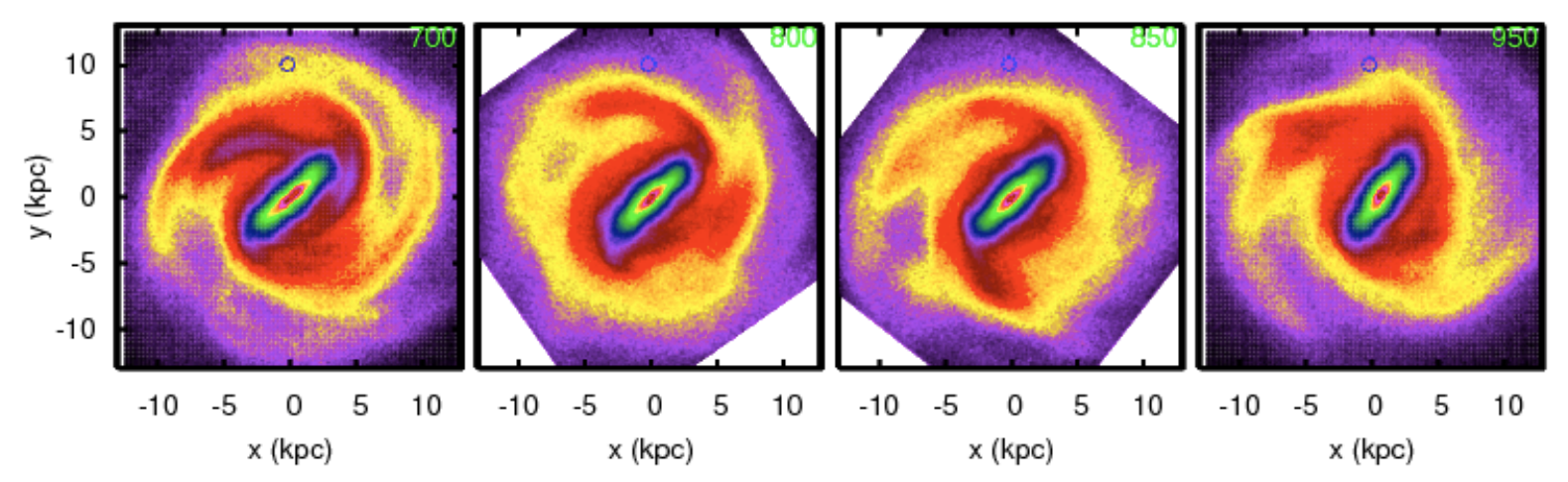} 
\caption{
For the four velocity distributions shown in Figure \ref{fig:r10_sn}
we show the morphology of the disk and the location (as small blue circles) 
of the neighborhoods
used to create these distributions.  The disk images have been flipped 
so that rotation is clockwise and rotated so that 
extraction neighborhoods are on the top. This has been done so that
these figures can more easily be compared to maps of the Galaxy.
The blue circle represents a neighborhood chosen to be like
the solar neighborhood.
As our simulated bar is longer than the Galactic one, the chosen
neighborhood is somewhat more distant than the Sun from the Galactic center.
The bar orientation is consistent with that inferred from infrared observations
of our Galaxy.  These images have in common a strong spiral image just
interior to the mock solar neighborhood.  This arm would
be consistent with the Centaurus Arm tangent arm.
\label{fig:sol}
}
\end{figure*}

Here we have only qualitatively attempted to relate spiral morphology
and patterns to structure seen in the velocity distributions.
However future work could aim to bridge the gap between test particle
simulations and  N-body simulations by understanding the orbits
in each feature seen in the velocity distributions.
Here we have focused on gross features present seen in the velocity
distributions, such as gaps and arcs.  Future studies
can also aim to understand the finer features as well, that are primarily
seen in smaller neighborhoods (e.g., see Figure \ref{fig:comp}).
These studies would help interpret forthcoming pencil
beam (e.g., \citealt{minchev08}) 
and large scale velocity and proper motion surveys of 
stars in the Galactic disk.

Here we have interpreted gaps in the velocity distributions in terms of changes in the shapes
of nearly closed orbits associated with different density waves.  However future studies could also
consider alternate interpretations for structure seen in velocity distributions 
such as transient spiral arms \citep{desimone04}, 
bifurcations in the families of periodic orbits \citep{katsanikas11}, 
chaotic orbits along spiral arms near a bar \citep{voglis06}, 
phase wrapping following an initial disturbance \citep{minchev09}
or individual Lindblad resonances  \citep{quillen05}.

\vskip 2.0truein
We thank Larry Widrow for giving us and helping us with his code
GalacticICS. 
We thank NVIDIA for the gift of four Quadro FX 5800 and two GeForce GTX 280 video cards.
We thank Evghenii Gaburov and Stefan Harfst
for making $phi$GRAPE and Sapporo available.
Support for this work was provided by NSF through award AST-0907841.

{}

\clearpage

\begin{table*}
\vbox to220mm{\vfil 
\caption{\large Estimated Pattern Speeds and Locations of Resonances \label{tab:tab1}}
\begin{tabular}{@{}lllcccccccc}
\hline
Pattern                & Time    &  Pattern speed& $r_{in}-r_{out}$ &ILR$_2$&ILR$_3$&ILR$_4$&CR   &OLR$_4$&OLR$_3$&OLR$_2$ \\
                       & (Myr)   & (radians/Myr) & (kpc)             & (kpc) & (kpc) & (kpc) &(kpc)&(kpc)  &(kpc)  &(kpc)   \\ \hline
Bar                    &350-950  & 0.040         &  0.5 --6          &       & 2.4   & 3.6   & 5.1 & 7.2   & 7.6   & 9      \\
Inner two-armed Spiral &350-950  & 0.030         &  3 --10           & 2.4   & 3.8   & 4.4   & 7.2 & 9.5   & 10    & 11     \\ 
Outer two-armed Spiral &350-950  & 0.015         &  6 --15           & 4.7   & 7.1   & 9.1   & 13.5& 17    & 18    & 19     \\ 
Three-armed Spiral     &350-950  & 0.023         &  3 --15           &       & 4.5   &       & 9.1 &       &  12.3 &        \\
\hline
Bar                    &750-1305 & 0.035         &  0.5 --6          &       & 3.6   & 3.0  & 6.0  & 8.1   &  9.1  &  10    \\
Inner two-armed spiral &750-1305 & 0.022         &  3 --10           &  3.2  & 4.9   & 6.0  & 9.5  & 12.0  & 13.3  &  14    \\
Outer two-armed Spiral &750-1305 & 0.013         &  6 --20           &  6.3  & 8.9   & 10.9 & 15   & 19    & 19.5  &  22    \\ 
Three armed Spiral     &750-1305 & 0.023         &  5 --15           &       & 4.5   &      & 9.1  &       & 12.3  &        \\
\hline
\end{tabular}
{\\
These pattern speeds are measured from features seen in the spectrograms 
shown in Figures \ref{fig:spec2} and \ref{fig:spec3}.  
The pattern speeds can be approximately converted to units of km s$^{-1}$~kpc$^{-1}$ by multiplying by 1000.
The times listed in the second column correspond to the range of times 
used to construct these spectrograms. 
The radii $r_{in}$ and $r_{out}$ are estimated inner and outer radii in kpc
of the individual patterns. 
Here ILR$_2$ and OLR$_2$ refer to the inner and outer $m=2$ Lindblad resonances.  
Likewise ILR$_3$ and ILR$_4$ refer to the 
the $m=3$ and $m=4$ inner Lindblad resonances.
When a resonance from one pattern is near that of another pattern,
one pattern could be coupled to or driving the other.
}
 \vfil}
\end{table*}


\begin{thebibliography}{}


\bibitem[Antoja et al.(2009)]{antoja09}
Antoja, T., Valenzuela, O., Pichardo, B., Moreno, E., 
Figueras, F., \& Fernandez, D. 2009, ApJ, 700, L78

\bibitem[Arifyanto \& Fuchs(2006)]{arifyanto06}
Arifyanto, M. I. \& Fuchs, B. 2006,  A\&A,  449, 533

\bibitem[Bekki \& Freeman(2003)]{bekki03}
Bekki, K., \& Freeman, K. C. 2003, MNRAS, 346, L11

\bibitem[Benjamin et al.(2005)]{benjamin05}
Benjamin, R. A., et al. 2005, ApJL, 630, L149

\bibitem[Binney \& Tremaine(1987)]{B+T}
Binney, J., \& Tremaine, S. 1987, Galactic Dynamics,
Princeton University Press, Princeton, NJ

\bibitem[Bovy(2010)]{bovy10}
Bovy, J. 2010,  ApJ, 725, 1676 

\bibitem[Chilingarian et al.(2010)]{chilingarian}
Chilingarian, I. V., Di Matteo, P.,  
Combes, F., Melchior, A.-L., \&  Semelin, B. 2010, A\&A, 518, 61

\bibitem[Chakrabarty \& Sideris(2008)]{dalia08}
Chakrabarty, D., \& Sideris, I. V. 2008, A\&A, 488, 161

\bibitem[Churchwell et al.(2009)]{churchwell09}
Churchwell, E. et al. 2009, PASP, 121, 213

\bibitem[Comparetta \& Quillen(2010)]{comparetta10}
Comparetta, J., \& Quillen, A. C. 2010, MNRAS in press,
 arXiv1005.4952

\bibitem[Contopoulos \& Grosbol(1986)]{cont86}
Contopoulos, G., \& Grosbol, P. 1986, A\&A, 155, 11

\bibitem[Contopoulos(1975)]{cont75}
Contopoulos, G.~1975, ApJ, 201, 566

\bibitem[Contopoulos(1988)]{cont88}
Contopoulos, G.~1988, A\&A, 201, 44

\bibitem[Dehnen \& Binney(1998)]{dehnen98}
Dehnen, W., \& Binney, J. J. 1998, MNRAS, 298, 387 

\bibitem[Dehnen(1999)]{dehnen99}
Dehnen, W. 1999, ApJ, 524, L35

\bibitem[Dehnen(2000)]{dehnen00}
Dehnen, W.  2000, AJ, 119, 800	

\bibitem[De Simone et al.(2004)]{desimone04}
De Simone, R., Wu, X., \& Tremaine, S. 2004, MNRAS, 350, 627	

\bibitem[Drimmel \& Spergel(2001)]{drimmel01}
Drimmel, R., \& Spergel, D. N.	2001, ApJ, 556, 181	

\bibitem[Eggen(1996)]{eggen}
Eggen, O.~J.~1996, AJ, 112, 1595

\bibitem[Famaey et al.(2004)]{famaey04}
Famaey, B., Jorrissen, A., Luri, X., Mayor, M.,
Udry, S., Dejonghe, H., \& Turon, C. 2005, A\&A,  430, 165

\bibitem[Famaey et al.(2008)]{famaey08}
Famaey, B., Siebert, A., \& Jorissen, A. 2008, A\&A,  483, 453	

\bibitem[Fux(2001)]{fux01}
Fux, R.~2001, A\&A 373, 511

\bibitem[Fux(2000)]{fux00}
Fux, R. 2000, 
Dynamics of Galaxies: from the Early Universe to the Present, 15th IAP meeting held in Paris, France, July 9-13, 1999, Eds.: Francoise Combes, Gary A. Mamon, and Vassilis Charmandaris ASP Conference Series, Vol. 197, ISBN: 1-58381-024-2, 2000, p. 27, also http://arxiv.org/abs/astro-ph/9910130

\bibitem[Gaburov et al.(2009)]{gaburov09}
Gaburov, E., Harfst, S., \& Portegies Zwart, S. 2009, New Astronomy, 14, 630

\bibitem[Gardner et al.(2010)]{gardner10}
Gardner, E., \& Flynn, C. 2010, MNRAS, 405, 545	

\bibitem[Gerhard(2010)]{gerhard10}
Gerhard, O. 2010,  
invited talk to appear in "Tumbling, twisting, and winding galaxies: Pattern speeds along the Hubble sequence", E. M. Corsini and V. P. Debattista (eds.), Memorie della Societa`Astronomica Italiana,
http://arxiv.org/abs/1003.2489 



\bibitem[Gomez et al.(2010)]{gomez10}
Gomez, F. A., Helmi, A., Brown, A. G. A., \& Li, Y.-S.
2010, MNRAS, 408, 935

\bibitem[Harfst et al.(2007)]{harfst07}
Harfst, S., Gualandris, A., Merritt, D., Spurzem, R., Portegies Zwart, S.,
\& Berczik, P.  2007,  New Astronomy, 12, 357.

\bibitem[Harsoula \& Kalapotharakos(2009)]{harsoula09}
Harsoula, M., \&  Kalapotharakos, C. 2009, MNRAS 394, 1605


\bibitem[Helmi \& deZeeuw(2000)]{helmi00}
Helmi, A., \& deZeeuw, P. T. 2000, MNRAS, 319, 657

\bibitem[Helmi et al.(2006)]{helmi06}
Helmi, A., Navarro, J. F., Nordstrom, B.,
Holmberg, J., Abadi, M. G., \& Steinmetz, M.
2006, MNRAS, 365, 1309

\bibitem[Henry et al.(2003)]{henry03}
Henry, A. L., Quillen, A. C., Gutermuth, R. 2003, AJ, 126, 2831

\bibitem[Holmberg et al.(2009)]{holmberg09}
Holmberg J., Nordstroem B., \& Andersen J.  2009, A\&A, 501, 941

\bibitem[Ideta(2002)]{ideta02}
Ideta, M.  2002, ApJ, 568, 190

\bibitem[Jog \& Combes(2009)]{jog09}
Jog, C. J., \& Combes, F. 2009, Physics Reports, 471, 75-111

 \bibitem[Katsanikas et al.(2011)]{katsanikas11}   
Katsanikas, M., Patsis, P. A., \& Pinotsis, A. D.	2011, MNRAS in press, arXiv1103.3981	

\bibitem[Kuijken \& Dubinski(1995)]{kuijken95}
Kuijken, K., \& Dubinski, J. 1995, MNRAS, 277, 1341

\bibitem[Levine et al.(2006)]{levine06}
Levine, E. S., Blitz, L., \& Heiles, C.	 2006, ApJ, 643, 881	
	
\bibitem[Lindblad(1926)]{lindblad26}
Lindblad, B. 1926, Ark. Mat. Astron. Fys. 19A, No. 27 
(Medd. Astron. Obs. Uppsala, No. 4)

\bibitem[Kalnajs(1979)]{kalnajs79}
Kalnajs, A. J. 1979, AJ, 84, 1697

\bibitem[Lepine et al.(2010)]{lepine10}
Lepine, J.R.D., Roman-Lopes, A., Abraham,  Z., Junqueira, T.C.  
Mishurov, Y. N.  2010, MNRAS, 477, in press,
http://arxiv.org/abs/arxiv.1010.1790 

\bibitem[Makino \& Aarseth(1992)]{makino92}
Makino, J., \& Aarseth, S. J. 1992, PASJ, 44, 141

\bibitem[Makino et al.(2003)]{makino03}
Makino, J., Fukushige, T., Koga, M., \& Namura, K. 2003, PASJ,  55, 1163

\bibitem[Masset \& Tagger(1997)]{masset97}
Masset, F., \& Tagger, M. 1997, A\&A , 322, 442 

\bibitem[Meidt et al.(2008)]{meidt08}
Meidt, S. E., Rand, R. J., Merrifield, M. R., Debattista, V. P.,  \&
Shen, J.  2008, ApJ, 676, 899	

\bibitem[Meidt et al.(2009)]{meidt09}
Meidt, S. E., Rand, R. J., \& Merrifield, M. R.
2009, ApJ, 702, 277


\bibitem[Meza et al.(2005)]{meza05}
Meza, A., Navarro, J. F., Abadi, M. G., \& Steinmetz, M.
2005, MNRAS, 359, 93

\bibitem[Minchev \& Quillen(2006)]{minchev06}
Minchev, I.,  \& Quillen, A. C. 
2006, MNRAS, 368, 623

\bibitem[Minchev \& Quillen(2008)]{minchev08}
Minchev, I., \& Quillen, A.~C.\ 2008, MNRAS, 386, 1579

\bibitem[Minchev et al.(2007)]{minchev07}
Minchev, I., Nordhaus, J., \& Quillen, A. C.	
 2007, ApJ, 664, L31	

\bibitem[Minchev et al.(2009)]{minchev09}
Minchev, I., Quillen, A. C., Williams, M., Freeman, K. C., Nordhaus, J., 
Siebert, A., \& Bienayme, O. 2009, MNRAS, 396, L56	

\bibitem[Minchev \& Famaey(2010)]{minchev10a}
Minchev, I.,  \& Famaey, B.  
2010, ApJ, 722, 112

\bibitem[Minchev et al.(2010b)]{minchev10b}
Minchev, I., Boily, C., Siebert, A., \& Bienayme, O.	
 2010, MNRAS, 407, 2122

\bibitem[Minchev et al.(2011)]{minchev11}
Minchev, I., Famaey, B., Combes, F., Di Matteo, P., 
Mouhcine, M., \& Wozniak, H. 2011, A\&A,  527, 147

\bibitem[Naoz \& Shaviv(2007)]{naoz07}
Naoz, S., \& Shaviv, N. J. 2007, New Astronomy, 12, 410

\bibitem[Nordstrom et al.(2004)]{nordstrom04}
Nordstrom, B., Mayor, M., Andersen, J., Holmberg, J., Pont, F.,
Jorgensen, B. R., Olsen, E. H., Udry, S., \&  Mowlavi, N.
2004, A\&A, 418, 989

\bibitem[Olling \& Dehnen(2003)]{olling03}	
Olling, R. P., \& Dehnen, W. 2003, ApJ, 599, 275	
	
\bibitem[Patsis \& Kaufmann(1999)]{patsis99}
Patsis, P. A., \& Kaufmann, D. E. 1999, A\&A, 352, 469

\bibitem[Quillen(2003)]{quillen03}
Quillen, A. C. 2003, AJ, 125, 785	

\bibitem[Quillen \& Minchev(2005)]{quillen05}
Quillen, A. C., \& Minchev, I. 2005, AJ, 130, 576	

\bibitem[Quillen et al.(2009)]{quillen09}
Quillen, A. C., Minchev, I., Bland-Hawthorn, J., \& Haywood, M.
 2009, MNRAS, 397, 1599	
 
\bibitem[Quinlan \& Tremaine(1992)]{quinlan92}
Quinlan, G. D., \& Tremaine, S. 1992, MNRAS, 259, 505

\bibitem[Rautiainen \& Salo(1999)]{rau99}
Rautiainen, P., \& Salo, H. 1999, A\&A, 348, 737

\bibitem[Revaz \& Pfenniger(2004)]{revaz04}
Revaz, Y., \& Pfenniger, D. 2004, A\&A, 425, 67

\bibitem[Saha et al.(2007)]{saha07}
Saha, K., Combes, F., \& Jog, C. J. 2007, MNRAS, 382, 419

\bibitem[Sellwood \& Sparke(1988)]{sellwood88}
Sellwood, J. A., \& Sparke, L. S. 1988, MNRAS, 231, 25	

\bibitem[Sellwood \& Lin(1989)]{sellwood89}
Sellwood, J. A., \& Lin, D. N. C. 1989, MNRAS, 240, 991 	

\bibitem[Shaviv(2003)]{shaviv03}
Shaviv, N. J.	2003, New Astronomy, 8, 39	

\bibitem[Shevchenko(2011)]{shev11}	
Shevchenko, I. I. 2011, ApJ, in press, 2010arXiv1012.3606S

\bibitem[Siebert et al.(2011)]{siebert11}
Siebert, A. et al. 2011, arXiv 1011.4092, accepted for publication in MNRAS

\bibitem[Sygnet et al.(1988)]{sygnet88}
Sygnet, J. F., Tagger, M., Athanassoula, E., \& Pellat, R. 1988, MNRAS, 232, 733

\bibitem[Tagger et al.(1987)]{tagger87}
Tagger, M., Sygnet, J. F., Athanassoula, E., \& Pellat, R.
1987, ApJ, 318, L43	

\bibitem[Vall\'ee (2008)]{vallee08}
Vall\'ee, J. P.	2008, AJ, 135, 1301

\bibitem[Vanhollebeke et al.(2009)]{vanhollebeke09}
Vanhollebeke E., Groenewegen M. A. T., \& Girardi L.  2009, A\&A, 498, 95


\bibitem[Voglis et al.(2006)]{voglis06}
Voglis, N. Stavropoulos, I., \& Kalapotharakos, C. 2006,  MNRAS, 372, 901

\bibitem[Widrow \& Dubinski(2005)]{widrow05}
Widrow, L. M., \& Dubinski, J. 2005, ApJ, 631, 838

\bibitem[Widrow et al.(2008)]{widrow08}
Widrow, L. M., Pym, B., \& Dubinski, J. 2008, ApJ, 679, 1239

\bibitem[Yuan \& Kuo(1997)]{yuan97}
Yuan, C., \& Kuo C. L.
1997, ApJ, 486, 750

\end{thebibliography}
\end{document}